\documentclass[%
reprint,
superscriptaddress,
showkeys,
 amsmath,amssymb,
pra,
floatfix,
]{revtex4-2}
\pdfoutput=1
\usepackage{graphicx}
\usepackage{dcolumn}
\usepackage{bm}
\usepackage{multirow}
\usepackage{tabularx}
\usepackage{array}
\newcolumntype{M}[1]{>{\centering\arraybackslash}m{#1}}
\newcolumntype{P}[1]{>{\centering\arraybackslash}p{#1}}
\usepackage{hyperref}

\begin{document}

\preprint{APS/123-QED}

\title{Unraveling the vector nature of generalized space-fractional Bessel beams}

\author{Aqsa Ehsan}
 \affiliation{%
NanoTech Lab, Department of Electrical Engineering, Information Technology University (ITU) of the Punjab Ferozepur Road, Lahore 54600, Pakistan\\
}

\author{Muhammad Qasim Mehmood}%
\affiliation{%
NanoTech Lab, Department of Electrical Engineering, Information Technology University (ITU) of the Punjab Ferozepur Road, Lahore 54600, Pakistan\\
}%

\author{Kashif Riaz}%
\affiliation{%
NanoTech Lab, Department of Electrical Engineering, Information Technology University (ITU) of the Punjab Ferozepur Road, Lahore 54600, Pakistan\\
}
\author{Yee Sin Ang}%
\email{yeesin$\_$ ang@sutd.edu.sg}
\affiliation{%
Singapore University of Technology and Design (SUTD), 8 Somapah Road, Singapore 487372, Singapore\\
}

\author{Muhammad Zubair}%
\email{muhammad.zubair@itu.edu.pk}

\affiliation{%
NanoTech Lab, Department of Electrical Engineering, Information Technology University (ITU) of the Punjab Ferozepur Road, Lahore 54600, Pakistan\\
}




\date{\today}

\begin{abstract}
We introduce an exact analytical solution of the homogeneous space-fractional Helmholtz equation in cylindrical coordinates. This solution, called vector Space-Fractional Bessel Beam (SFBB), has been established from the Lorenz' gauge condition and Hertz vector transformations. We perform scalar and vector wave analysis focusing on electromagnetics applications, especially in cases where the dimensions of the beam are comparable to its wavelength $(k_r \approx k)$. The propagation characteristics such as the diffraction and self-healing properties have been explored with particular emphasis on the polarization states and transverse propagation modes. Due to continuous order orbital angular momentum dependence, this beam can serve as a bridge between the ordinary integer Bessel beam and the fractional Bessel beam and, thus, can be considered as a generalized solution of the space-fractional wave equation that is applicable in both integer and fractional dimensional spaces. The proposed SFBBs provide better control over the beam characteristics and can be readily generated using Digital Micromirror Devices (DMDs), Spatial Light Modulators (SLMs), metasurfaces, or spiral phase plates. Our findings offer new insights on electromagnetic wave propagation, thus paving a new route towards novel applications in optical tweezers, refractive index sensing, optical trapping, and optical communications.
\end{abstract}

\maketitle


\section{\label{Intro}Introduction}
Localized light beams, such as Airy beams, Vortex beams, Bessel beams and Mathieu beams have always been intriguing because of their remarkable focusing properties. Localized light beams are represented by the solution of scalar wave equation or the homogeneous Helmholtz equation in free space. They have considerable applications in almost every scientific and technological field such as optical tractor beam, optical trapping, optical communications, and the photolithography of nanoscale transistors where a focused light beam or a high intensity gradient is required \cite{structured1, structured2}. 
\par
Among all these structured light beams, Bessel beams are of particular importance owing to their exceptional self-reconstructing and diffraction-free propagation. Bessel beams were first introduced in the late $20^{th}$ century by Brittingham and others who came up with the solutions of the Helmholtz equation in terms of cylindrical coordinates \cite{miller1977symmetry,yu1}. Such beams were initially named the Focus Wave Modes (FWM) as they had a focused intensity profile throughout their propagation. Later, Durnin and Dholakia came up with the exact representations of zero and higher order Bessel beams \cite{yu5durnin1,yu7}. It was also shown that the higher order Bessel beams possess orbital angular momentum (OAM) and can be used to trap and manipulate particles \cite{yu8,opticalMicromanipulation,OpticaltrappingCounter-propagating}. Similarly, Lynn showed that high intensity optical tweezers can be made using these Beams which can work successfully in the micrometer and nanometer range \cite{LynnThesis}.These optical tweezers have vast range of applications in medical field and in laser fabrication. 

\par
Bessel beams also hold strong potential in communication technology. They are capable of transferring several terabits of data at a time without being affected by the turbulence present in free space \cite{adaptiveFreeSpaceOpticalCoomunication,Terabitfree-space,Du:15}.In quantum regime, Bessel beams have been employed to reduce the losses due to entangled photon pairs during quantum communication in free space \cite{quantumEntanglement}. These applications make Bessel beams a strong candidate for the future communication systems. 

\par
Another advantage of Bessel beams is their ease of generation. They can be easily generated using axicons, metasurfaces, holograms, spiral phase plates and spatial light modulators (SLMs) \cite{CapassoBessel, YangSpiral:18, wangbessel1,wangbessel2}. Highly efficient Bessel beams have been generated using the aforementioned techniques \cite{HighefficiencyBesselbeamarraygenerationbyHuygensmetasurfaces,highlyEfficientBessel,unravelingbesselbeams}. This ease of generation makes them suitable for various applications such as on-chip optical tweezers and laser cutters, but due to their integer order orbital angular momentum they have limited bandwidth and their spot size cannot be precisely controlled especially during nano-fabrication. To address this issue Tao introduced fractional order OAM in the general representation of the Bessel beams \cite{Tao1,Tao2,Tao3,Tao4}. It was also shown experimentally that these beams can be used in particle manipulation. But, because of their fractional definition of OAM, they cannot be considered as a true analytical solution of the wave equation \cite{TaoComment}. Vega introduced another representation of the fractional Bessel beams as a linear combination of multiple integer Bessel beams \cite{Vega1,Vega2}. Such a solution satisfies the homogeneous Helmholtz equation but its scalar nature restricts its use to only those cases where the dimensions of the beams are much greater than its wavelength $(k_r >> k)$. Other solutions include Mitri's and Caloz's vector Bessel beams defined in the Cartesian and Cylindrical coordinate systems respectively \cite{Mitri1,Mitri2,Mitri3,caloz}. Both of these solutions were the vector versions of Vega's solutions and thus satisfy the Helmholtz equation but due to the superposition of multiple beams of different orders, these beams possess non-vortex nature. Another approach towards fractionalization is the fractional-dimensional approach as introduced by Zubair et al. \cite{ZubairBook,ZubairWavePropagation,ZubairWaveEquation,ZubairNovelFractional,zubair4,zubair5,zubair14,zubair12,exactCylindrical,frac_dimension}.  Using the same approach we have derived an exact analytical solution of the space-fractional cylindrical wave equation regarded here as the space-fractional Bessel beams (SFBBs). The initial results of this analysis were presented elsewhere \cite{SFBB}. This approach not only helps in achieving higher bandwidth and better control over the beam's spot size but it also helps in modeling the beam's behavior in the presence of fractal geometries. Moreover, like other fractional and integer Bessel beams, and structured light beams, the proposed SFBBs can also be generated easily using SLMs that have the ability to modulate the amplitude, phase or polarization of the light waves in space and time and are widely used now a days to generate arbitrary shaped beams such as anti-diffracting optical pin beams, fractional Bessel beams, vortex beams and optical tweezers etc. at the desired wavelengths with high efficiency \cite{SLM1,SLM2,SLM3}.
\par
In this paper, complete scalar and vector analysis of the space-fractional Bessel beams will be discussed. Section \ref{scalarAnalysis} describes the scalar analysis of the SFBB and Section \ref{vectorAnalysis} contains complete vector analysis. Scalar analysis will be helpful in determining the propagation characteristics such as the intensity and phase profiles of the beam during propagation and the effect of different disturbances on the beam's profile, and vector analysis helps in determining the power and energy, and the transverse modes of the beam, especially, when the central defect of the beam $k_r$ becomes comparable to its wavelength $k$. A detailed comparison of space-fractional Bessel beams with other fractional Bessel beams is shown in Section \ref{comparison} and the prospective applications of these beams are briefly discussed in Section \ref{applications}. 

\section{Full-Wave Analysis of the Space-Fractional Bessel Beams}
\subsection{\label{scalarAnalysis}Scalar Analysis}
Free space scalar wave equation for the space-fractional electric and magnetic field components can be written as
\begin{equation}\label{eq:1}
    \nabla_D^2 \textbf{E} + k^2 \textbf{E} = 0
\end{equation}
\begin{equation}\label{eq:2}
    \nabla_D^2 \textbf{H} + k^2 \textbf{H} = 0
\end{equation}
where, $\textbf{E}$ and $\textbf{H}$ represent the electric and magnetic fields respectively, and, $k =\frac{2\pi}{\lambda}$ is the wave number. Time dependency is suppressed during the whole discussion. $\nabla_D ^2 $ or Laplacian operator used above can be defined for cylindrical coordinate system in fractional-dimensional space as \cite{SFBB}

\begin{eqnarray}\label{eq:3}
    \nabla_D^2 ={}& \frac{\partial^2}{\partial \rho^2} \space +  \frac{(\alpha_1+\alpha_2-1)}{\rho}\frac{\partial}{\partial \rho} + \frac{1}{\rho^2}\bigg(\frac{\partial^2}{\partial \phi^2}-\nonumber\\&\{(\alpha_1-1)\tan\phi - (\alpha_2-1)\cot\phi\}\frac{\partial}{\partial\phi}\bigg) +     \\& \frac{\partial^2}{\partial z^2}+\frac{\alpha_3-1}{z}\frac{\partial}{\partial z}.\nonumber
\end{eqnarray}

$\rho$, $\phi$ and $ z $ in the above equation represent each coordinate of the cylindrical coordinate system. Other parameters representing the spatial dimensions are $0\leq\alpha_1\leq1$, $0\leq\alpha_2\leq1$ and $0\leq\alpha_3\leq1$. Each one of these parameters represents each cylindrical coordinate such that the overall dimension of the space ($D=\alpha_1+\alpha_2+\alpha_3$) becomes three as in the case of ordinary integer dimensional space. It is important to note here that each of these coordinates acts independently and can thus be tailored according to the requirements. In order to derive the expression for scalar SFBB we will be solving the second order partial differential equation in Eq. (\ref{eq:1}). It should be noted that the solution of Eq. (\ref{eq:2}) will be similar to Eq. (\ref{eq:1}) due to duality.
The electric field in terms of cylindrical coordinate system is expressed as  

\begin{eqnarray}\label{eq:4}
    \textbf{E}(\rho,\phi,z)={}& E_\rho(\rho,\phi,z)\boldsymbol{\hat{\rho}}+ E_\phi(\rho,\phi,z)\boldsymbol{\hat{\phi}}+\\& E_z(\rho,\phi,z)\boldsymbol{\hat{z}}\nonumber,
\end{eqnarray}

where, $\boldsymbol{\hat{\rho}}$, $\boldsymbol{\hat{\phi}}$ and $\boldsymbol{\hat{z}}$ represent unit vectors in the cylindrical coordinate system. After substituting the expression of $ \textbf{E}(\rho,\phi,z)$ in Eq. (\ref{eq:1}), and then performing separation of variables, we get,
\begin{equation}\label{eq:5}
    \bigg[\rho^2\frac{d^2}{d \rho^2} \space +  (\alpha_1+\alpha_2-1){\rho}\frac{d}{d \rho}+(k_\rho\rho)^2-m^2\bigg]f(\rho)=0,
\end{equation}
\begin{eqnarray}\label{eq:6}
    \bigg[{}&\frac{d^2}{d \phi^2}-\{(\alpha_1-1)\tan\phi - (\alpha_2-1)\cot\phi\}\frac{d}{d\phi}+ \\&m^2\bigg]g(\phi)=0\nonumber,
\end{eqnarray}
\begin{equation}\label{eq:7}
    \bigg[\frac{d^2 }{d z^2}+\frac{\alpha_3-1}{z}\frac{d}{dz} + k^2_z\bigg]h(z)=0,
\end{equation}
where, 
\begin{equation}\label{eq:8}
    k^2_\rho+k^2_z=k^2.
\end{equation}
Solving for $\rho$, $\phi$ and $z$, and after combining their final solutions we get \cite{Abramowitz,Stillinger},
\begin{equation}\label{eq:9}
\begin{split}
     \psi(\rho,\phi,z)={}& \rho^{1-\frac{\left(\alpha_1+\alpha_2\right)}{2}}\sin^{2-\alpha_2}\phi \cos^{2-\alpha_1}\phi\times\\&\bigg[C_1 J_{v_1} \left(k_\rho\rho\right)+C_2 Y_{v_1} \left(k_\rho\rho\right)\bigg] \times \\&  \bigg[C_3F\left(\alpha,\beta,\gamma;\zeta\right)+  C_4\zeta^{1-\gamma}\\&F\left(\alpha-\gamma+1,\beta-\gamma+1,2-\gamma;\zeta \right)\bigg] \\&  \times e^{kz}U(a,b,v_2),
\end{split}   
\end{equation}
where,
\begin{equation*}
   v_1=\frac{1}{2}\sqrt{(2-\alpha_1-\alpha_2)^2+4m^2} 
\end{equation*}
is the OAM dependent order of the Bessel functions $J_{v_1}$ and $Y_{v_1}$ of first and second kind respectively,  $m$ represents the geometrical charge or the OAM, $F$ denotes the Gaussian hypergeometric function. Other parameters are
\begin{equation*}
     \alpha+\beta+1=\frac{1}{2}\left(8-\alpha_2-\alpha_1\right),
\end{equation*}
\begin{equation*}
    \alpha\beta=\frac{1}{4}\left(8-2\alpha_1-2\alpha_2-m^2\right),
\end{equation*}
\begin{equation*}
    \gamma=\frac{1}{2}\left(4-\alpha_2\right), k=\frac{\sqrt{D}-a_1}{2a_2},
\end{equation*}
\begin{equation*}
    D=a_1^2-4a_0a_2, v_2=\frac{z-\mu}{\lambda}, \mu=\frac{-b_2}{a_2},
\end{equation*}
\begin{equation*}
    \lambda=\frac{-a_2}{2a_2k+a_1}, a=\frac{B(k)}{2a_2k+a_1},
\end{equation*}
\begin{equation*}
    b=\frac{a_2b_1-a_1b_2}{a_2^2}, B(k)=b_2k^2+b_1k+b_0,
\end{equation*}
and, $U(a,b,v_2)$ is the Kummer Hypergeometric function.
\par
\begin{figure}[ht!]
\centering
\includegraphics[width=1\columnwidth, keepaspectratio]{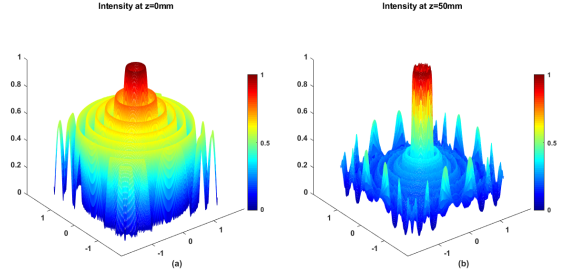}

\caption{Intensity profiles of s-SFBB at (a) $0$ $mm$ and (b) $50$ $mm$. $\alpha_1, \alpha_2$ and $\alpha_3$ are set to $0.75, 0.75$ and $1$ respectively such that the overall dimension $D=2.5$}
\label{fig:IntensityAt0and100mm}
\end{figure}

Eq. (\ref{eq:9}) will now be termed as the scalar-space-fractional Bessel beam (s-SFBB). Here, it should be kept in mind that the order of this s-SFBB is not a direct function of the geometrical topological charge rather it is a function of $v_1$ and for a particular value of $m$, depending upon the spatial dimensions, $v_1$ can have multiple fractional values. The three dimensional intensity profile of the s-SFBB is shown in Fig. \ref{fig:IntensityAt0and100mm}. Here, the intensity profile is shown at two different lengths along the  longitudinal or the z-axis. The spatial dimension of the beam is $D=2.5$. Propagation results are obtained using the angular spectrum approach. For the theoretical generation, the overall grid size of $3.84$ $mm$ by $3.84$ $mm$ is considered which is further subdivided into 256 pixels in each dimension. Working wavelength is set to $633$ $nm$ and the radial wave number $k_\rho=13$ $mm^{-1}$. It is assumed that the beam is fractional only along the transverse plane ($\rho$ and $\phi$ plane, $\alpha_3=1$) and thus behaves in the same way along the z-axis as the ordinary integer Bessel beam. 

\par
\begin{figure}[ht!]
\centering
\includegraphics[width=1\columnwidth, keepaspectratio]{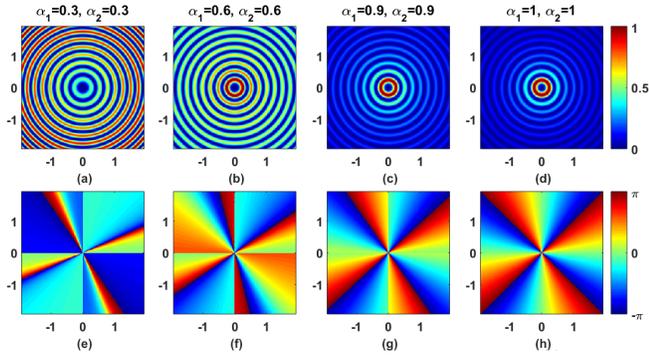}

\caption{Two-dimensional intensity and phase plots of the s-SFBB for multiple values of $\alpha_1$ and $\alpha_2$ with $\alpha_3=1$}
\label{fig:IntensitysameAlpha}
\end{figure}
\begin{figure}[htb!]
\centering
\includegraphics[width=1\columnwidth, keepaspectratio]{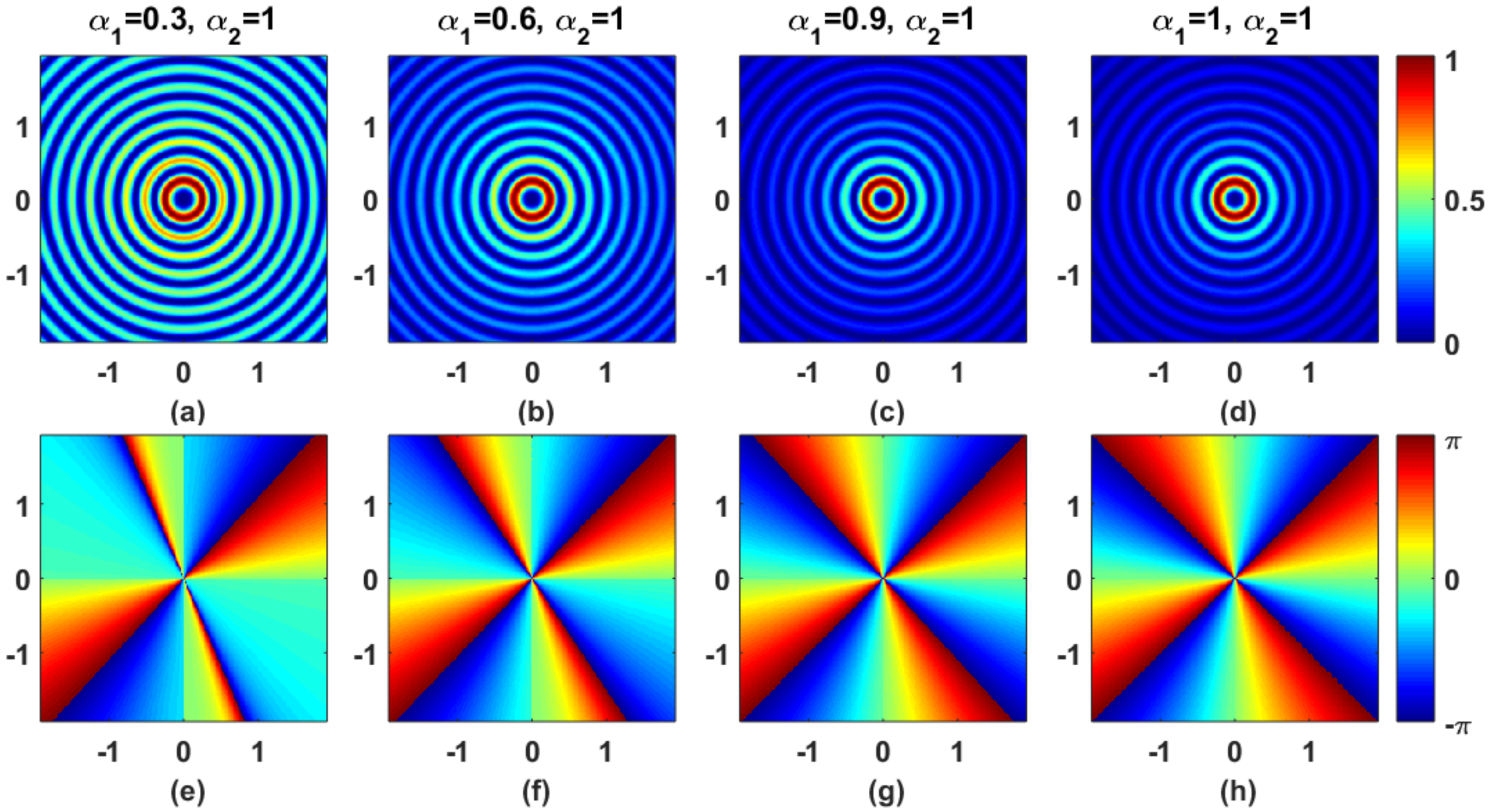}

\caption{Two-dimensional intensity and phase plots of the s-SFBB for multiple values of $\alpha_1$ with $\alpha_2=\alpha_3=1$}
\label{fig:IntensityDifferentAlpha1}
\end{figure}
\begin{figure}[htb!]
\centering
\includegraphics[width=1\columnwidth, keepaspectratio]{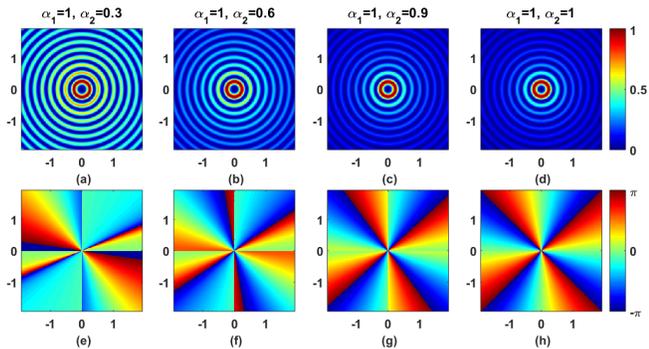}

\caption{Two-dimensional intensity and phase plots of the s-SFBB for multiple values of $\alpha_2$ with $\alpha_1=\alpha_3=1$}
\label{fig:IntensityDifferentAlpha2}
\end{figure}

Two-dimensional intensity and phase plots of the s-SFBB are shown in Fig. \ref{fig:IntensitysameAlpha}. Here, $\alpha_1$ and $\alpha_2$ are varied equally such that the total dimension $1\leq D \leq 3$. Moreover, it can be seen that as the dimension is increased from $1.6$ to $3$, more and more energy flows from the outer rings towards the inner ones such that when the dimension becomes $D=3$, almost all of the energy of the beam is concentrated in the two innermost rings. While considering phase of this beam it can be observed that there are jumps or discontinuities in the phase depicting the vortex nature of the beam. This aspect of the beam will prove handy in the applications involving optical manipulation of the particles. A similar situation is observed when only $\alpha_1$ or $\alpha_2$ is varied keeping the other two dimensions constant. This behaviour is depicted in Fig. \ref{fig:IntensityDifferentAlpha1} where only $\alpha_1$ is varied and in Fig. \ref{fig:IntensityDifferentAlpha2} where $\alpha_2$ is varied. All other parameters remain unchanged. Hence, it can be concluded that for fractional dimensions, the intensity of the beam distributes itself among the outer rings and for integer orders it is concentrated in the inner rings. Furthermore, it possesses vortex nature due to the existence of orbital angular momentum or the geometrical topological charge, and thus can be classified as a vortex beam. 

\par
\begin{figure}[htb!]
\centering
\includegraphics[width=0.75\columnwidth, keepaspectratio]{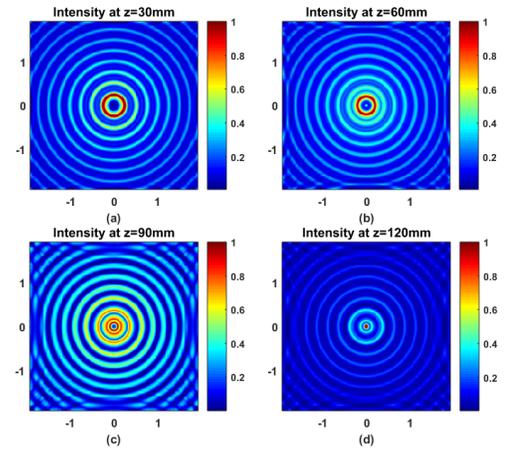}

\caption{Two-dimensional intensity plots of the s-SFBB at multiple distances along the z-axis for $\alpha_1=0.75$, $\alpha_2=0.75$ and $\alpha_3=1$}
\label{fig:PropagationMultipleDistances}
\end{figure}

The scalar wave theory also helps in determining the propagation characteristics of the beam while it propagates in the free space. This can be done by making use of the angular spectrum approach according to which the intensity of the beam at a distance $z$ along the propagation axis can be calculated by taking the inverse Fourier transform of its angular spectrum. Mathematically,
\begin{equation}\label{angularSpectrum}
    \psi(\rho,\phi,z)=f\{\hat{f}\left[\psi(\rho,\phi,z-\Delta z)\right]exp(-ik_z \Delta z)\},
\end{equation}
where, $\hat{f}$ and $f$ are used to denote the Fourier and Inverse Fourier Transforms respectively, $z$ represents the generation distance, and $\Delta z$ is the point of observation. Once the intensity profile at a given distance is calculated, it can be compared with the intensity of an integer Bessel Beam to verify that whether the beam is diffraction-free or not. This analysis is shown in Fig. \ref{fig:PropagationMultipleDistances} where it can be observed that as the distance increases or as the beam propagates along the free space, it maintains its intensity distribution proving that the s-SFBB is diffraction-free in nature.

\par
\begin{figure}[t!]
\centering
\includegraphics[width=1\columnwidth, keepaspectratio]{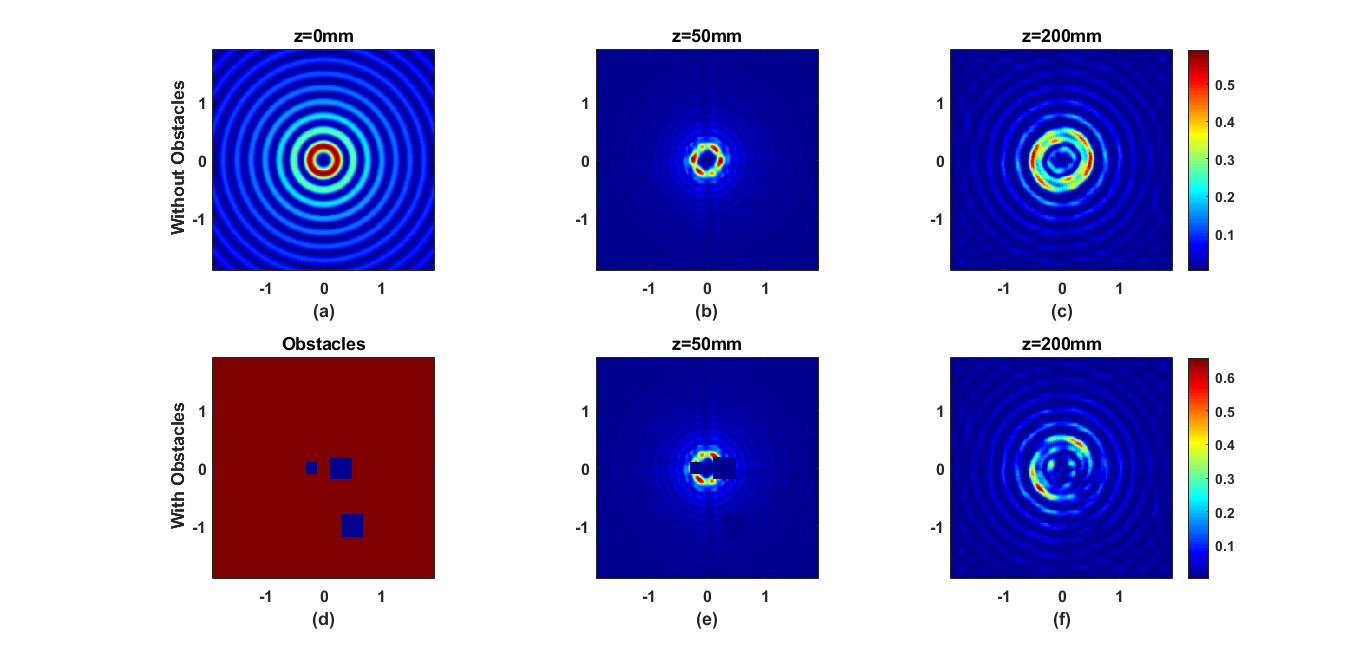}

\caption{Two-dimensional intensity plots of the s-SFBB at multiple distances with and without obstacles (a) Intensity at $0$ $mm$ (b),(c) Intensity at $50$ $mm$ and $200$ $mm$ in the absence of obstacles (d) Obstacles placed at $50$ $mm$ (e),(f) Intensity at $50$ $mm$ and $200$ $mm$ in the presence of obstacles. Overall sizes of the obstacles are $0.4$ $mm$ $\times$ $0.4$ $mm$ (right and bottom) and $0.2$ $mm$ $\times$ $0.2$ $mm$ (left).}
\label{fig:FrationalWithAndWithoutObstacles}
\end{figure}
\begin{figure}[htbp!]
\centering
\includegraphics[width=1\columnwidth, keepaspectratio]{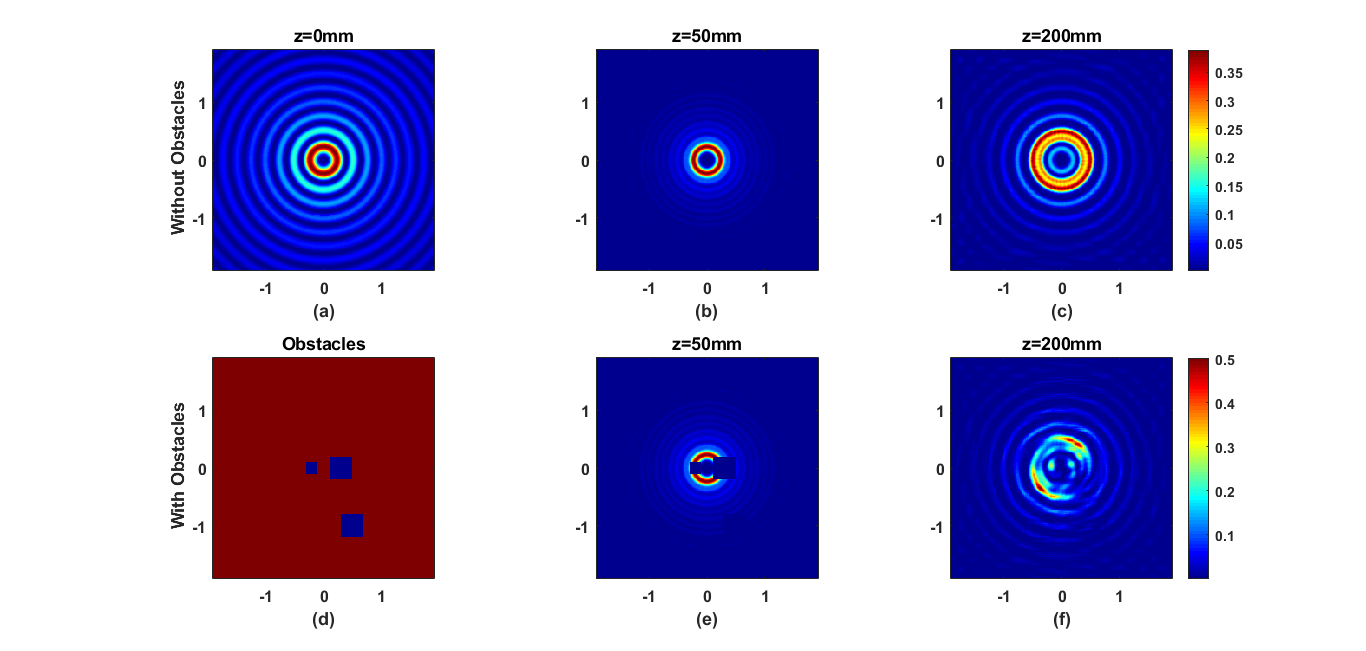}

\caption{Two-dimensional intensity plots of the ordinary integer Bessel beam at multiple distances with and without obstacles (a) Intensity at $0$ $mm$ (b),(c) Intensity at $50$ $mm$ and $200$ $mm$ in the absence of obstacles (d) Obstacles placed at $50$ $mm$ (e),(f) Intensity at $50$ $mm$ and $200$ $mm$ in the presence of obstacles. Overall sizes of the obstacles are $0.4$ $mm$ $\times$ $0.4$ $mm$ (right and bottom) and $0.2$ $mm$ $\times$ $0.2$ $mm$ (left).}
\label{fig:IntegerWithAndWithoutObstacles}
\end{figure}
\begin{figure}[htbp!]
\centering
\includegraphics[width=1\columnwidth, keepaspectratio]{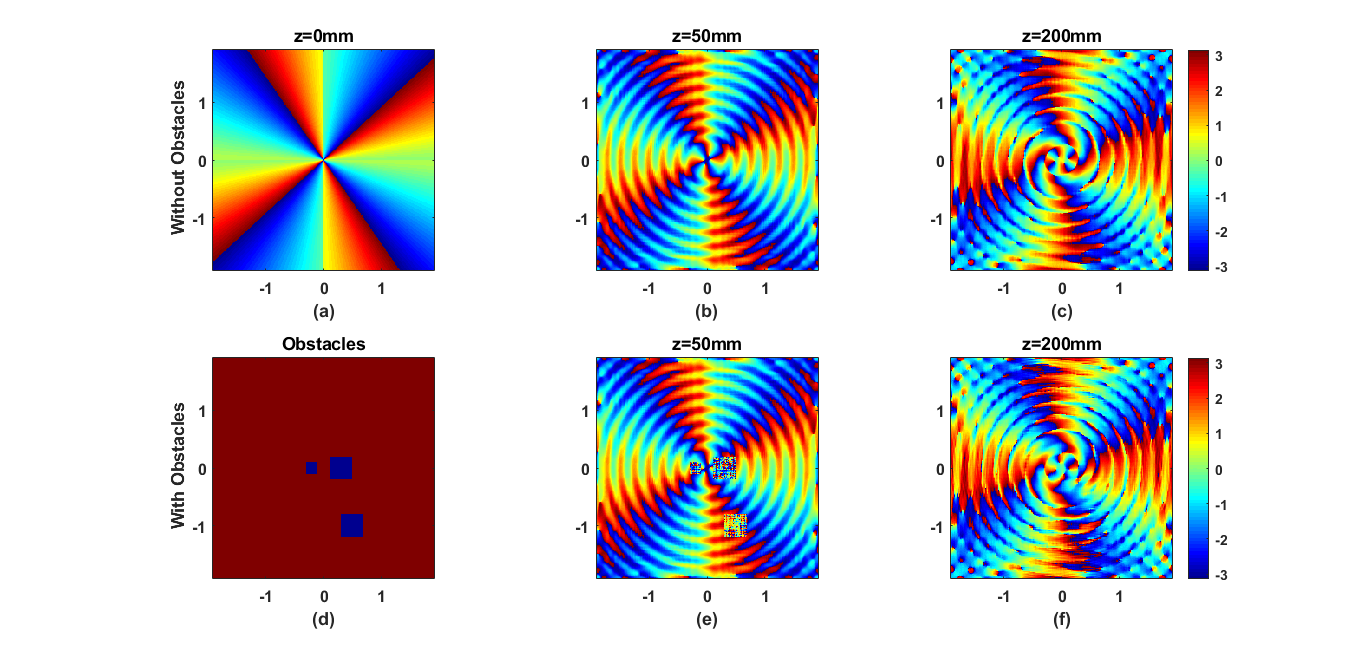}

\caption{Two-dimensional phase plots of the s-SFBB at multiple distances with and without obstacles (a) Phase at $0$ $mm$ (b),(c) Phase at $50$ $mm$ and $200$ $mm$ in the absence of obstacles (d) Obstacles placed at $50$ $mm$ (e),(f) Phase at $50$ $mm$ and $200$ $mm$ in the presence of obstacles. Overall sizes of the obstacles are $0.4$ $mm$ $\times$ $0.4$ $mm$ (right and bottom) and $0.2$ $mm$ $\times$ $0.2$ $mm$ (left).}
\label{fig:PhaseFractionalWithAndWithoutObstacles}
\end{figure}
\begin{figure}[htbp!]
\centering
\includegraphics[width=1\columnwidth, keepaspectratio]{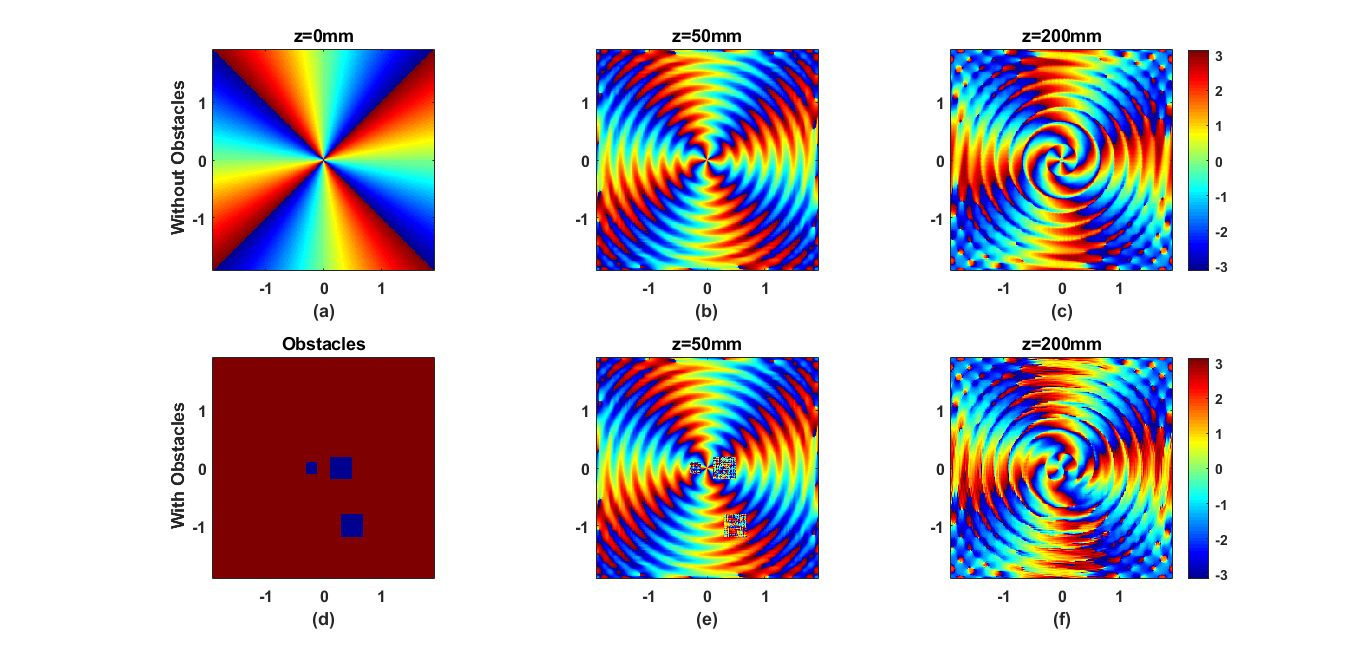}

\caption{Two-dimensional phase plots of the ordinary integer Bessel beam at multiple distances with and without obstacles (a) Phase at $0$ $mm$ (b),(c) Phase at $50$ $mm$ and $200$ $mm$ in the absence of obstacles (d) Obstacles placed at $50$ $mm$ (e),(f) Phase at $50$ $mm$ and $200$ $mm$ in the presence of obstacles. Overall sizes of the obstacles are $0.4$ $mm$ $\times$ $0.4$ $mm$ (right and bottom) and $0.2$ $mm$ $\times$ $0.2$ $mm$ (left).}
\label{fig:PhaseIntegerWithAndWithoutObstacles}
\end{figure}

This approach can also be used to prove the self-reconstruction property of the s-SFBB. For this purpose, we placed three obstacles in the path of the beam at a distance of $50$ $mm$ and observed the intensity profile of the beam at $200$ $mm$. This profile was then compared with the intensity profile of this beam in the absence of obstacles as shown in Fig. \ref{fig:FrationalWithAndWithoutObstacles}. Same analysis was done with the Integer Bessel beams as shown in Fig. \ref{fig:IntegerWithAndWithoutObstacles} and the results show that this beam can reconstruct itself after experiencing disturbances. It was also observed that lower intensity rings can be recovered easily when obstructed but it is difficult to recover higher intensity rings as more amount of energy is required to fill up these rings. This also depends upon the size of the obstacles. It is easier to recover from small sized obstacles as compared to the larger ones. Another factor that affects reconstruction of the beam is the placement of the obstacle. Obstacles placed in the dark or low energy areas of the beams do not affect the intensities much and vice versa. In the same manner, the effect of obstacles on the phase of the s-SFBB and the ordinary Integer Bessel beam was observed as shown in Figs. \ref{fig:PhaseFractionalWithAndWithoutObstacles} and (\ref{fig:PhaseIntegerWithAndWithoutObstacles}) respectively. Both of these results were in accordance with the previously discussed observations. Hence, it can be concluded that the s-SFBB is non-diffracting and self-healing in nature.   

\subsection{\label{vectorAnalysis}Vector Analysis}
Scalar space-fractional Bessel beam can be transformed into vector representation by making use of the Hertz vector potentials and the Lorenz Gauge condition as follows,
\begin{equation}\label{eq:10}
    \boldsymbol{E_e}=\boldsymbol{\nabla_D}(\boldsymbol{\nabla_D}\cdot\boldsymbol{\Pi_e})-k^2\boldsymbol{\Pi_e}
\end{equation}
\begin{equation}\label{eq:11}
    \boldsymbol{H_e}=-ik\frac{1}{\eta}(\boldsymbol{\nabla_D}\times\boldsymbol{\Pi_e})
\end{equation}
\begin{equation}\label{eq:12}
    \boldsymbol{E_m}=ik\eta(\boldsymbol{\nabla_D}\times\boldsymbol{\Pi_m})
\end{equation}
\begin{equation}\label{eq:13}
    \boldsymbol{H_m}=\boldsymbol{\nabla_D}(\boldsymbol{\nabla_D}.\boldsymbol{\Pi_m})-k^2\boldsymbol{\Pi_m}
\end{equation}
where,  $e$ and $m$ in the subscript are used to denote the TMn and TEn modes respectively and the total field can be calculated by adding both of these modes, $k=\frac{2\pi}{\lambda}$ is the wave number, $\boldsymbol{\nabla_D}$ is same as defined in Eq. (\ref{eq:3}), and, $\boldsymbol{\Pi_e}$ and $\boldsymbol{\Pi}_m$ are the corresponding Hertz vector potentials which can be expressed using the equation mentioned below,
\begin{eqnarray}\label{eq:14}
     \boldsymbol{\Pi_{e/m}}={}&\Pi_{e/m} \boldsymbol{\hat{z}}=A_{e/m}\rho^{1-\frac{(\alpha_1+\alpha_2)}{2}}\bigg[C_1J_{v_1}(k_\rho\rho)\bigg]\times\nonumber\\& \sin^{2-\alpha_2}\phi \cos^{2-\alpha_1}\phi \times
     \bigg[C_3F(\alpha,\beta,\gamma;\zeta)+\\&C_4\zeta^{1-\gamma}F(\alpha-\gamma+1,\beta-\gamma+1,2-\gamma;\zeta)\bigg]\times\nonumber\\&e^{kz}U(a,b;v_2)\boldsymbol{\hat{z}}\nonumber,
\end{eqnarray}
Here, $\Pi_{e/m}$ is the scalar electric/magnetic potential and it can be expressed in a similar way as the s-SFBB. Hankel functions and time dependency is suppressed in this case for simplicity. Detailed analytical expressions of the transverse modes of SFBB are discussed in the supplementary material. Fig.~\ref{fig:EH_TM0} shows the intensity graphs of TMn mode of the vector-Space Fractional Bessel Beam (v-SFBB). The wavelength is set to $633$ $nm$ as in the case of s-SFBB, $\alpha_1$ and $\alpha_2$ are equal while $\alpha_3=1$ in all cases. All other parameters are same as that of the s-SFBB. Here, it can be seen that in case of integer order of the beam the circular symmetry exists between the intensity profiles but for fractional dimensions this symmetry is broken. Thus, for fractional dimensional space, circular symmetry in the intensity profiles of the v-SFBB will not exist.  

\begin{figure*}[htb!]
\centering
\includegraphics[width=1.8\columnwidth, keepaspectratio]{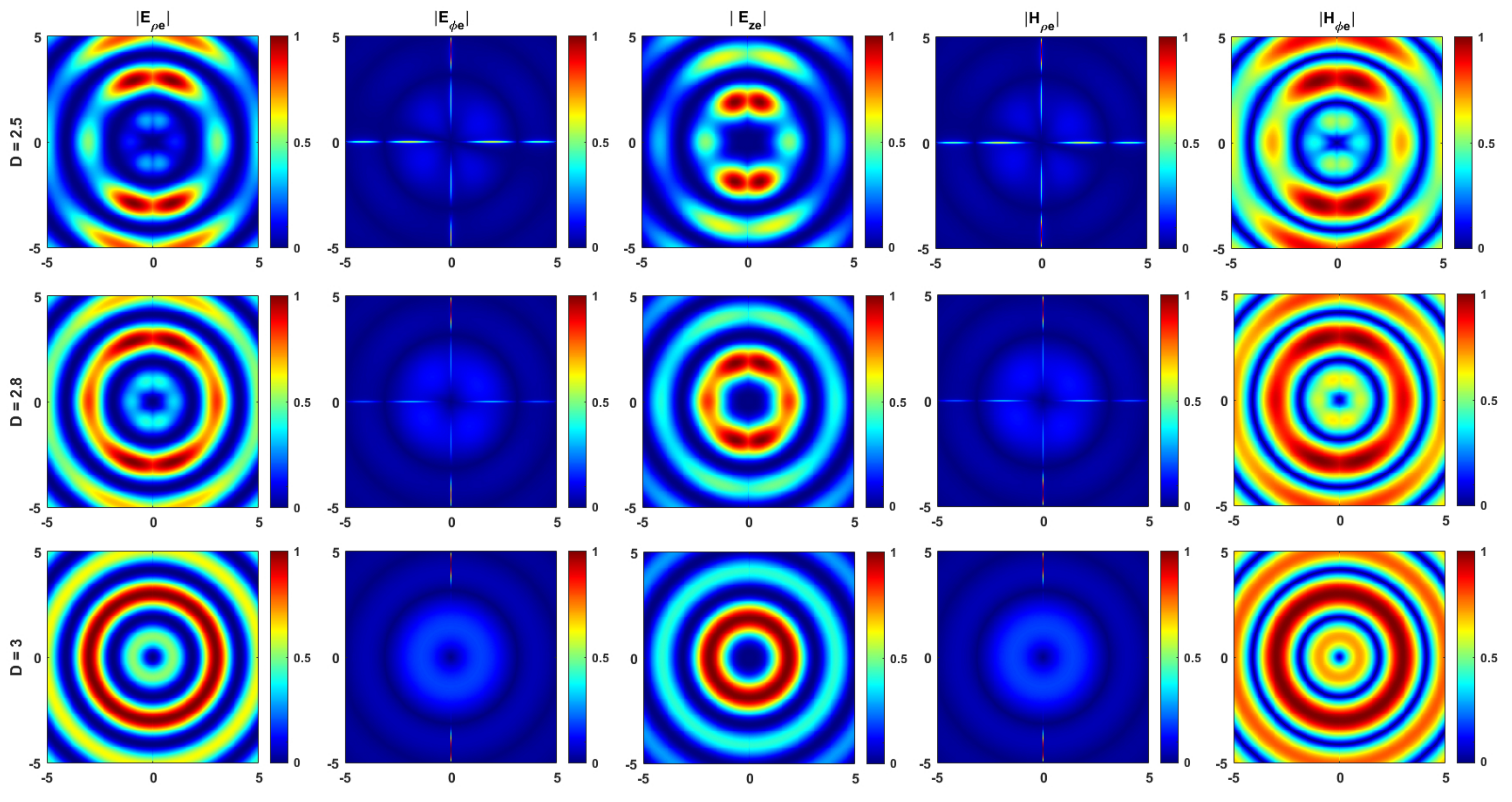}

\caption{Transverse magnetic modes of the v-SFBB for different spatial dimensions}
\label{fig:EH_TM0}
\end{figure*}

\par
\begin{figure}[htbp!]
\centering
\includegraphics[width=1\columnwidth, keepaspectratio]{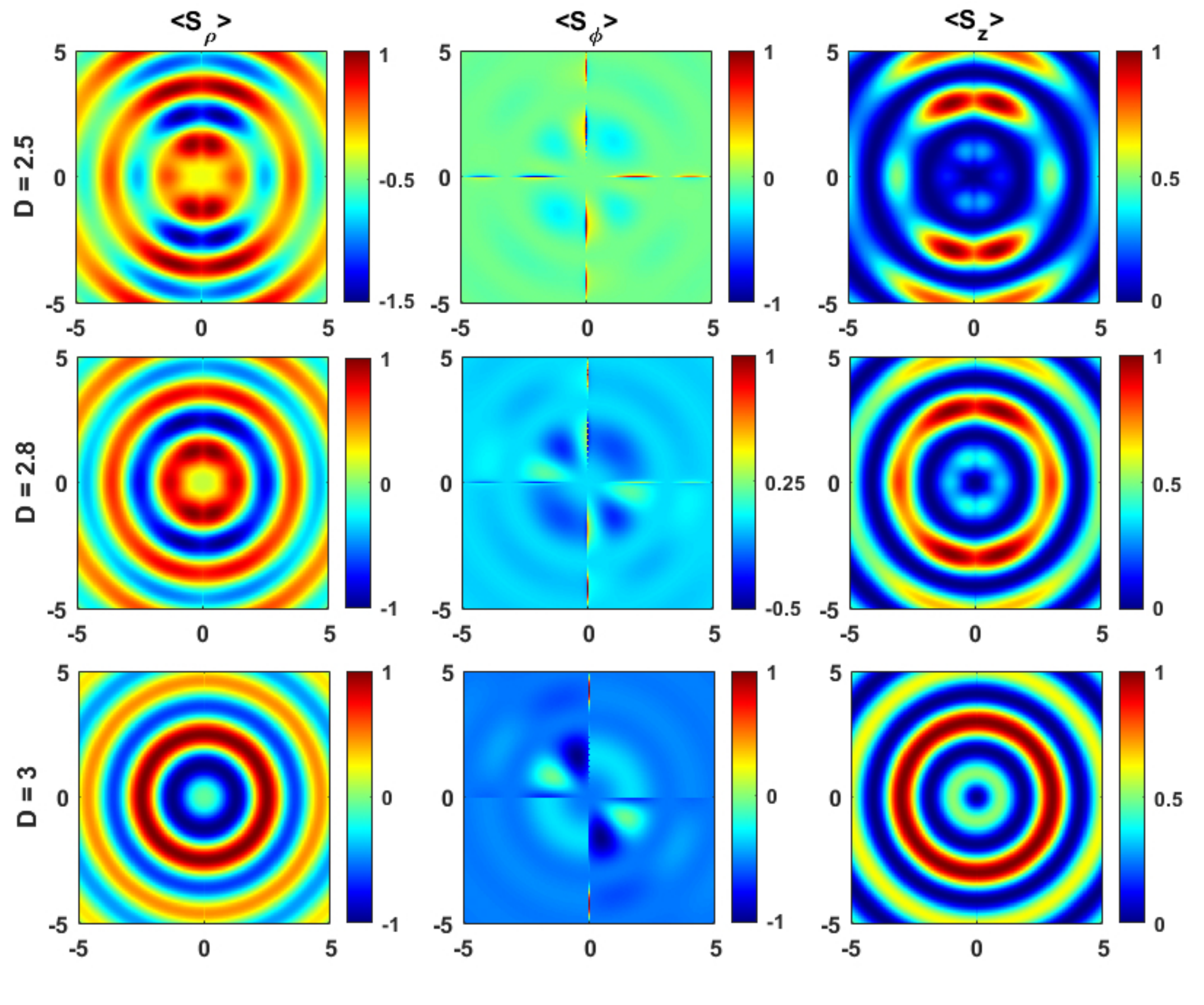}

\caption{Poynting vector plots of the v-SFBB for different spatial dimensions}
\label{fig:Poynting_EH_TM0}
\end{figure}

The time-averaged Poynting vector for the v-SFBB can be derived using the expression given below,
\begin{equation}\label{eq:43}
    \boldsymbol{<S>}=\frac{1}{2}Re(\boldsymbol{E \times H^*}).
\end{equation}
$\boldsymbol{E}$ and $\boldsymbol{H}$ in the above expression are used to denote the complex electric and magnetic fields respectively. Fig.~\ref{fig:Poynting_EH_TM0} shows the Poynting vector plots of the v-SFBB. A comparison with the integer counterpart is also shown in the same figure. Here, it can be seen that the average power $<S_z>$ vanishes at the beam axis so that the energy flow along the propagation direction $(z)$ is distributed over the rings. While comparing the intensities of both integer and v-SFBBs, it can be concluded that the circular symmetry does not exist for fractional order orbital angular momentum. This property can be utilized in devices such as spin jets to separate out two or more particles from a mixture as it creates a strong intensity gradient within the rings of the SFBB. It can also be employed in refractive index sensors for liquids.   

\section{\label{comparison}Comparing Space-Fractional Bessel Beams with Existing Fractional Bessel Beams}
Fractional Bessel Beams were first introduced by Tao et al \cite{Tao1,Tao2,Tao3,Tao4}. Their proposed solution was the same as that of an integer Bessel Beam except that it had fractional amplitude and phase components owing to fractional order orbital angular momentum. He and his colleagues showed that this type of beam can travel through obstacles placed in its path and it has the ability to reconstruct itself. They also showed that the experimental generation of such a beam is easily possible by the help of a digital micromirror device or a programmable spatial light modulator and it has the ability to trap and manipulate micro-meter sized particles. But this type of solution does not retain the non-diffractive property of the integer Bessel beam as it showed an opening when propagated in space. Also due to its scalar nature, properties like TE and TM modes and polarization etc. cannot be determined. Furthermore, it does not satisfy the homogeneous Helmholtz equation because this expression satisfies the scalar wave equation only for integer order orbital angular momentum \cite{TaoComment}. This drawback of Tao's solution was resolved by Vega et al. \cite{Vega1,Vega2}. He showed that a non-integer solution of the homogeneous Helmholtz equation can be obtained by linearly adding up the integer order Bessel beams. Such a solution is an exact solution as it satisfies the closure property and can be reduced to the integer Bessel beam by setting the fractional operator $\alpha$ equal to some integer value. Vega also showed experimental generation of such beams by the help of Computer Generated Holograms and proved that these beams do not diffract while propagating in space. But due to the scalar nature of his proposed solution, it cannot be used in electromagnetics, as its propagation characteristics cannot be studied completely without deriving its vector solution because in the case of electromagnetics the radial spot or the dimensions of the beam are comparable to its wavelength and thus vector analysis cannot be ignored. To encounter this issue Mitri et al. proposed a vector solution using Vega's scalar potential in terms of Cartesian  coordinates \cite{Mitri1,Mitri2,Mitri3}. This solution was diffraction-free in nature but it lacked circular symmetry. Caloz came up with the solution to this problem by introducing Vector Bessel beams using Vega's solution in terms of cylindrical coordinates \cite{caloz}. Our proposed solution is different from the Fractional Bessel beams discussed above as it is defined using the  fractional-dimensional approach. This exact analytical solution not only satisfies the free space scalar wave equation but is also self-reconstructing and diffraction-free in nature. This is depicted in Fig. \ref{fig:FrationalWithAndWithoutObstacles} where the intensity profiles of the s-SFBB are observed in the presence of obstacles and are compared with the intensity profiles in case of no obstacles. Similar analysis is done with the phase in Fig. \ref{fig:PhaseFractionalWithAndWithoutObstacles} where the s-SFBB is propagated in space using the angular spectrum approach and obstacles are placed in its path to observe its behaviour after passing through an obstruction. Here, it can be seen that this beam can easily reconstruct itself after experiencing an obstruction and is also diffraction-free in nature. Comparison of the intensity plots of s-SFBB with the ordinary integer Bessel beam under the presence of obstacles can be observed in Fig. \ref{fig:IntegerWithAndWithoutObstacles}. Similarly, phase comparison can be seen in Fig. \ref{fig:PhaseIntegerWithAndWithoutObstacles}. All of these results are summarized in Table \ref{table1}.

 \begin{table*}
 \small
\renewcommand{\arraystretch}{1}
\caption{A brief comparison of reported fractional Bessel beams}
\label{table1}

\begin{ruledtabular}
\begin{tabular}{ P{1cm} P{2cm} P{2.5cm} P{3 cm} P{2 cm} P{2cm}   }

{\textbf{Ref.}} & {\textbf{Diffraction Property}} & {\textbf{Self Healing Property  }} & {\textbf{Fractional Parameter }}& {\textbf{Nature of the solution }}& {\textbf{Type of Solution }}\\
\hline
\textbf{\cite{Tao1}}  & Diffracting & Self-Healing & Fractional order OAM $(m)$ & Scalar & Approximate \\

\textbf{\cite{Vega1}} & Non-Diffracting & --- & Superposition of Integer Bessel Beams (fractional $\alpha$) & Scalar & Exact \\

\textbf{\cite{Mitri1}} & Non-Diffracting & --- & Superposition of Integer Bessel Beams (fractional $\alpha$) & Vector  & Exact \\

\textbf{\cite{caloz}} & Non-Diffracting & --- & Superposition of Integer Bessel Beams (fractional $\alpha$) & Vector & Semi-Analytical \\

\textbf{\cite{SFBB}} & Non-Diffracting & Self-Healing & Fractional Dimensional Space $(D=\alpha_1+\alpha_2+\alpha_3)$ & Scalar  & Exact \\

\textbf{This Work} & Non-Diffracting & Self-Healing & Fractional Dimensional Space $(D=\alpha_1+\alpha_2+\alpha_3)$ & Vector  & Exact \\
\end{tabular}
\end{ruledtabular}
\end{table*}
 

\section{\label{applications} Prospective Applications of Space-Fractional Bessel Beams (SFBBs)}
Our proposed v-SFBB can have remarkable applications in the fields of optical engineering and optical communications such as optical tweezers, spin jets, refractive index sensors and optical multiplexers. In this section, we discuss some of the possible applications of these beams in detail. 
\subsection{Optical Trapping and Manipulation}
A structured light beam is said to possess orbital angular momentum if it exhibits azimuthal phase dependence of $e^{jm\phi}$ where $m$ is the geometrical topological charge. Such a beam possesses an orbital angular momentum equal to $m\hbar$ per photon and a phase of $2\pi m$. The mathematical expression of the phase term of our s-SFBB suggests that it carries OAM. This is also evident in the intensity plots of the phase profile of our s-SFBB. The OAM possessed by our proposed SFBB relates to its spatial dimensions according to the following expression,
\begin{equation}\label{eq:oam}
    m=\frac{1}{2}\sqrt{4v_1^2-(2-\alpha_1-\alpha_2)^2}
\end{equation}
Here, it must be kept in mind that the phase of this beam is not directly dependent on $m$ rather it is dependent on $v_1$ and for a particular $m$, depending upon the spatial dimensions of the beam, we can achieve multiple values of $v_1$. Another interesting point is that, the variation of $v_1$ with $m$ decreases as the value of $m$ increases such that for higher values of $m$, it is approximately of the same value. This variation of $v_1$ with $m$ gives complete control over the OAM and the order of the beam. So, just by playing with  the dimensions of the beam, we can easily achieve any value of the OAM. This can be used to generate both vortex and non-vortex beams by carefully adjusting the dimensions of the beam and by superimposing two or more beams. Also, the continuous orbital angular momentum dependence on the dimensions of the beam plays an important role in creating intensity gradients within the beam to carefully manipulate particles placed in its path. This aspect of SFBB and other Bessel beams in general helps in accurate positioning of the trapped particles in comparison with other structured light beams such as Gaussian beams where it is difficult to precisely trap or place particle at the desired location. Furthermore, it can also prove handy in the fabrication of sub-wavelength sized metamaterials and metasurfaces due to its focused intensity profile and non-diffracting behavior along the propagation axis.


\par
\begin{figure}[htbp!]
\centering
\includegraphics[width=1\columnwidth, keepaspectratio]{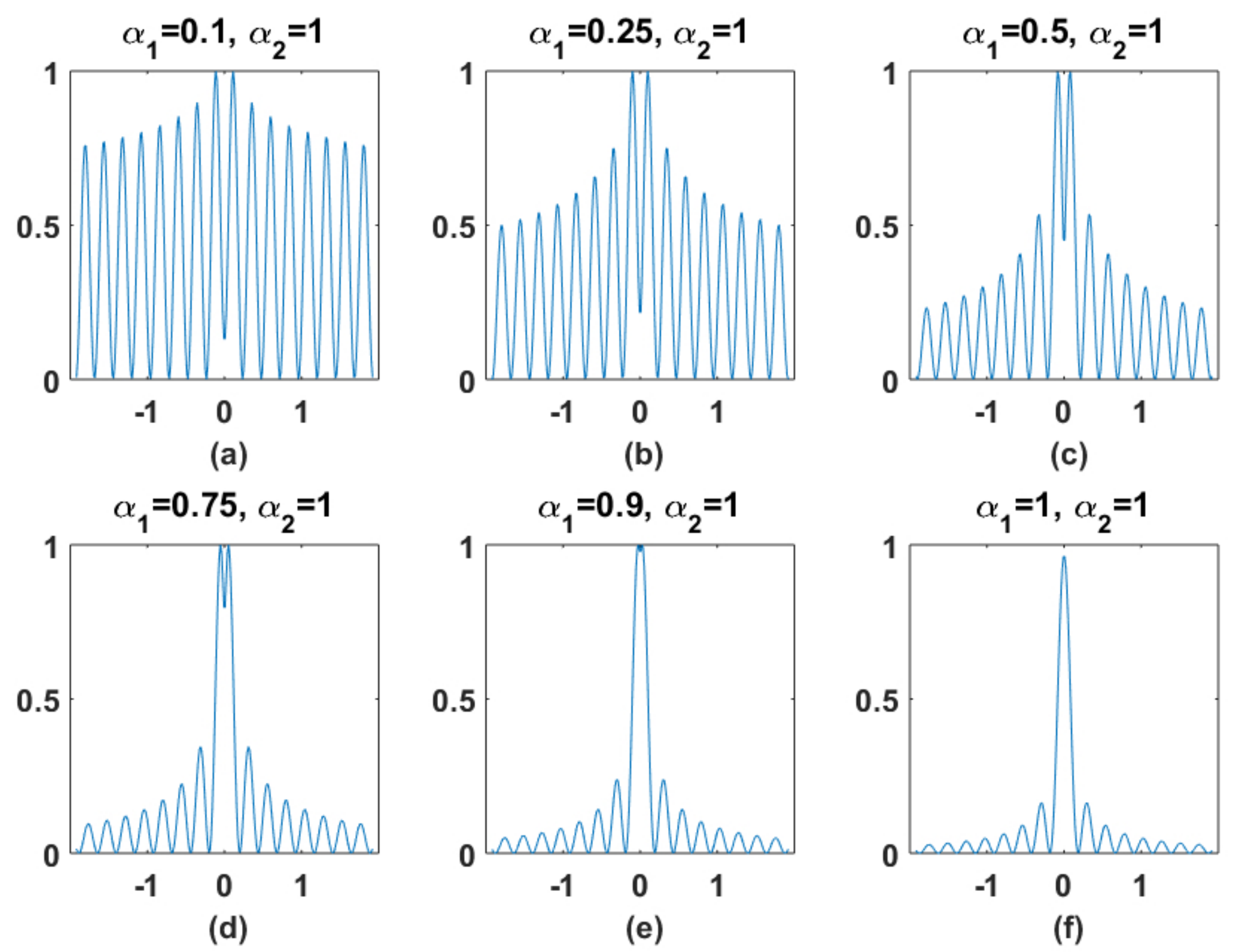}

\caption{1D intensity graphs showing FWHMs of s-SFBB for different values of $\alpha_1$ keeping $\alpha_2=\alpha_3=1$ }
\label{fig:varyingAlpha1}
\end{figure}
\begin{figure}[htbp!]
\centering
\includegraphics[width=1\columnwidth, keepaspectratio]{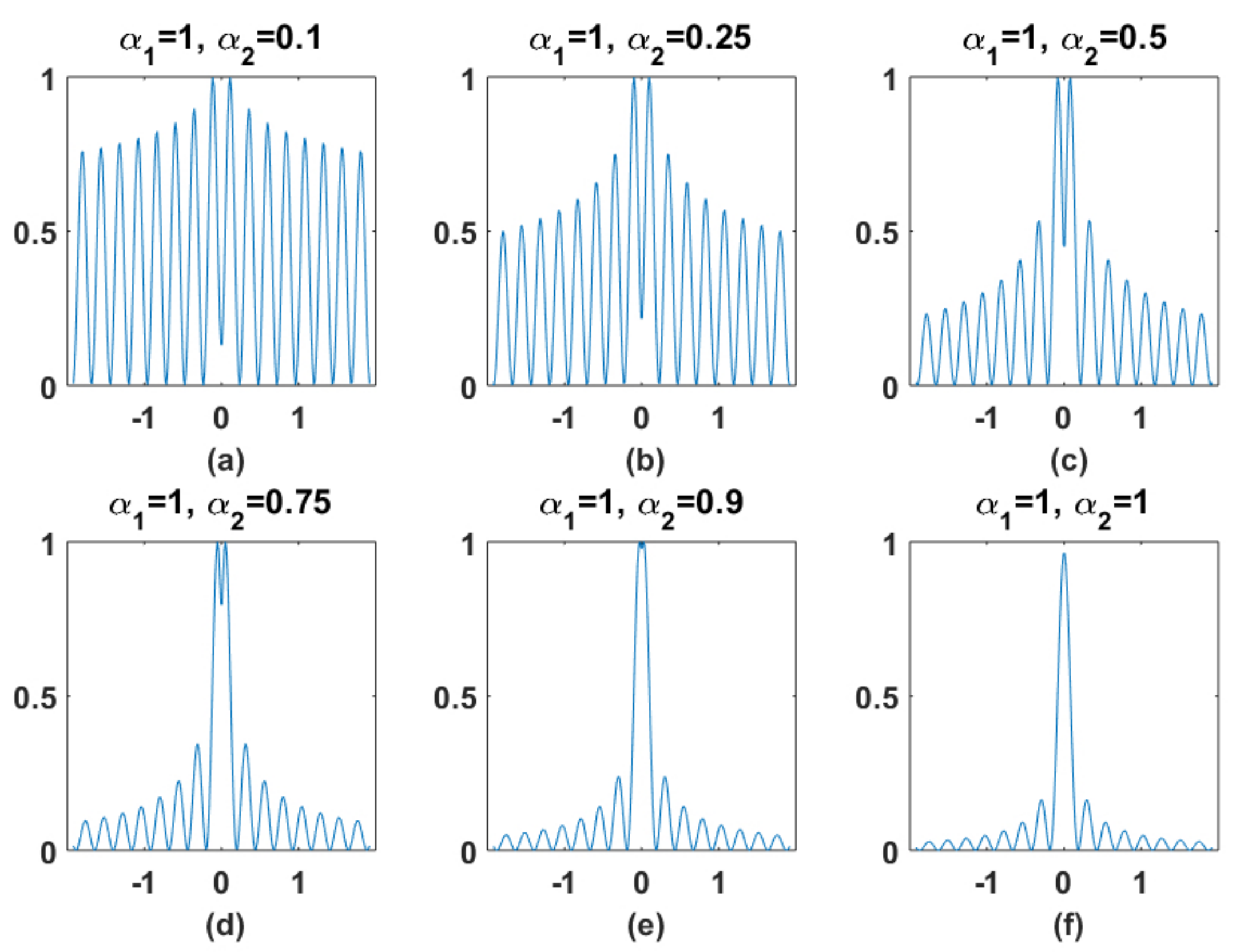}

\caption{1D intensity graphs showing FWHMs of s-SFBB for different values of $\alpha_2$ keeping $\alpha_1=\alpha_3=1$ }
\label{fig:varyingAlpha2}
\end{figure}
\begin{figure}[htbp!]
\centering
\includegraphics[width=1\columnwidth, keepaspectratio]{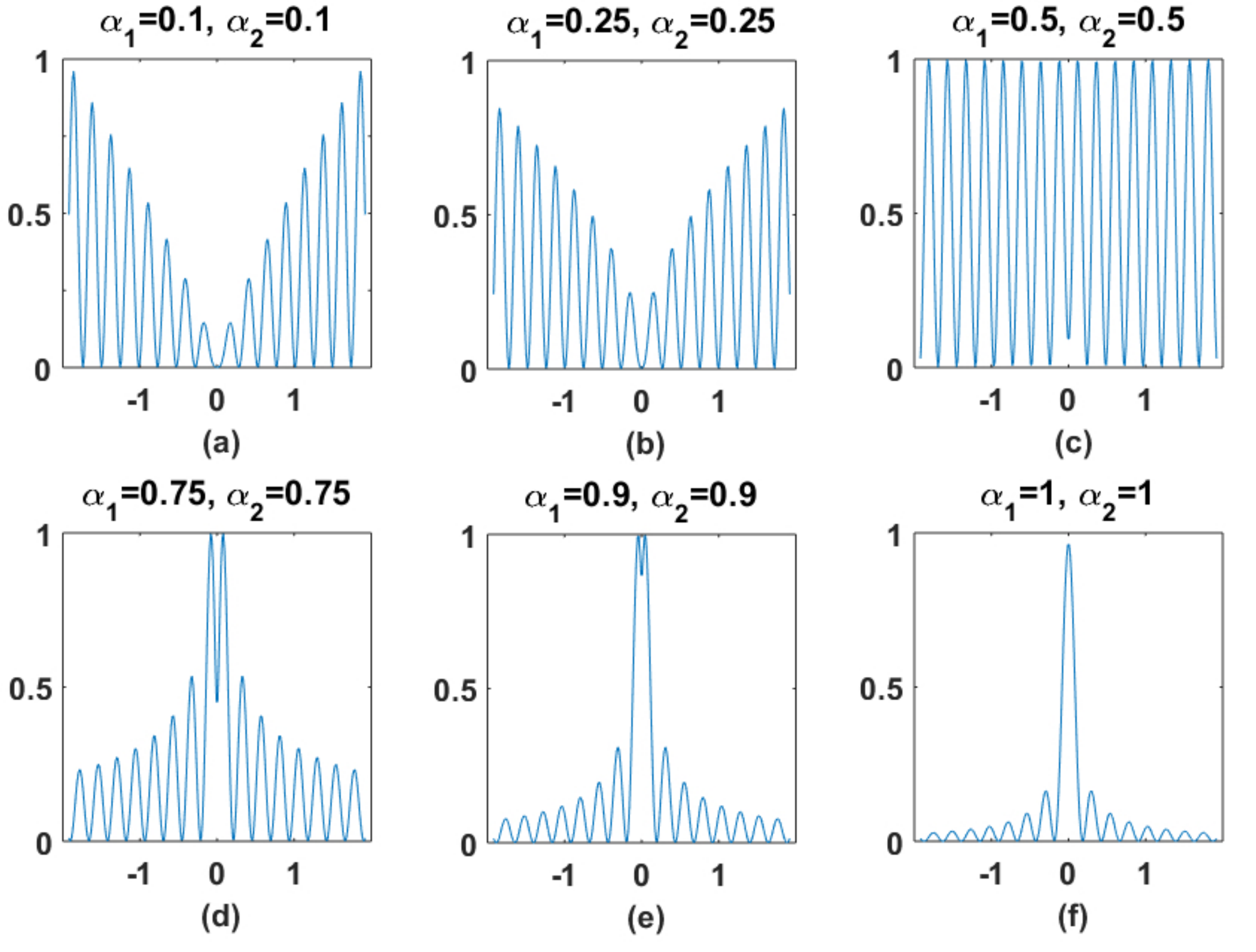}

\caption{1D intensity graphs showing FWHMs of s-SFBB for different values of $\alpha_1$ and $\alpha_2$ keeping $\alpha_3=1$ }
\label{fig:varyingBoth}
\end{figure}

 As depicted in Figs. \ref{fig:varyingAlpha1},  \ref{fig:varyingAlpha2} and \ref{fig:varyingBoth} the v-SFBB beam can have different values of Full-Width Half Maximums depending upon the dimensions of the beam. This aspect along with the OAM perspective can be used to easily trap nanometer and micrometer sized particles just by optimizing the dimensions of the beam.
 
\subsection{Optical Communications}
\begin{figure}[htbp!]
\centering
\includegraphics[width=1\columnwidth, keepaspectratio]{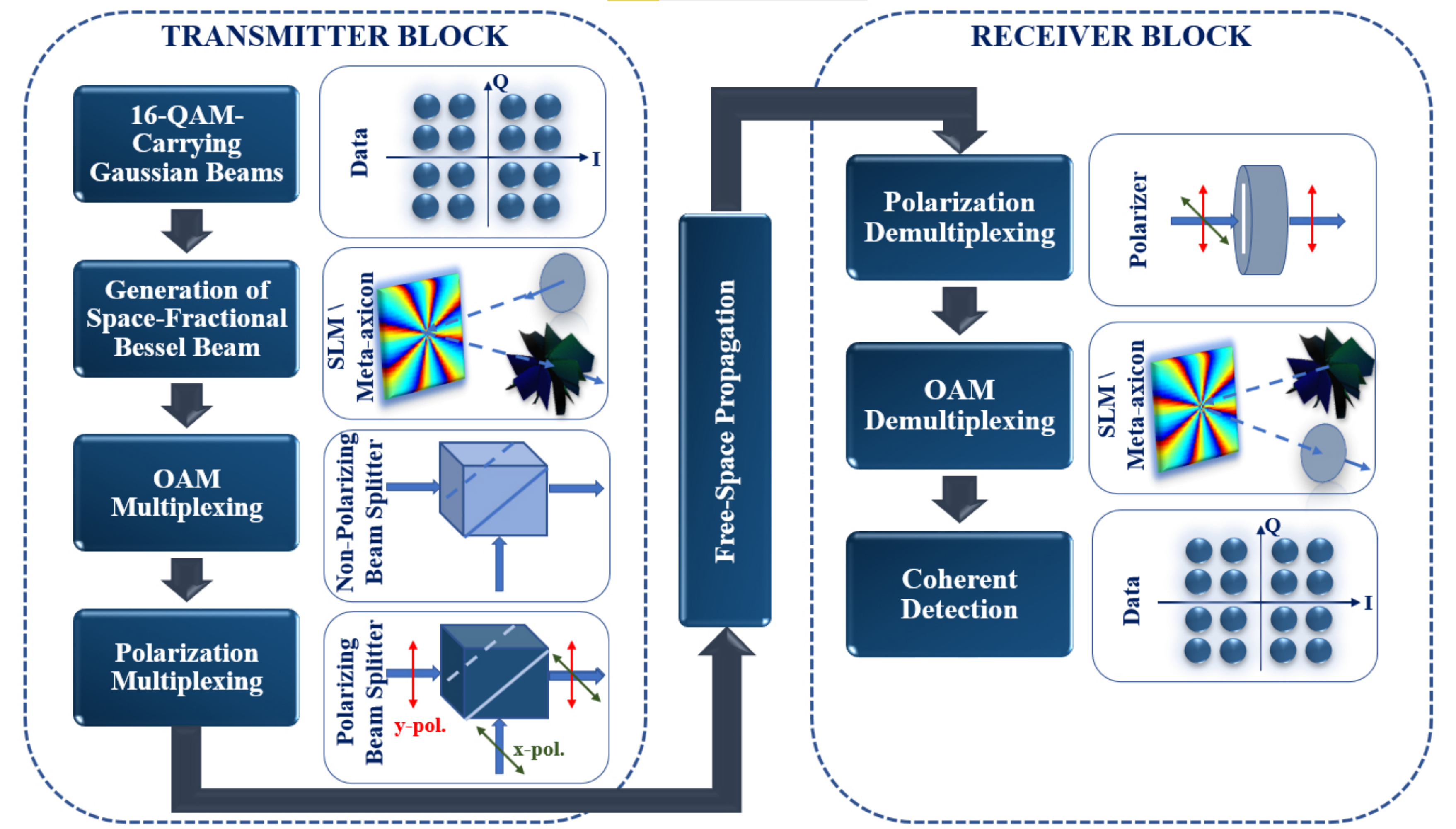}

\caption{Proposed experimental setup for OAM multiplexing/ demultiplexing using SFBBs}
\label{fig:opticalCommunications}
\end{figure}

Another interesting application linked to the OAM perspective of this beam is the OAM multiplexing for communication in the air or free space. The proposed experimental setup for OAM multiplexing/ demultiplexing through free space is shown in Fig. \ref{fig:opticalCommunications}. Large amounts of data can be encoded within the intensity rings using different orders of the geometrical topological charge of the beam with minimum bit error rate and can be transferred easily to a distant location as this OAM multiplexing not only reduces the cross-talk but also shows resilience to the turbulence effects present in the free space. The data rate can further be increased by increasing the value of $m$ but this too has its own consequences. Also, the performance of such devices depends upon the accurate generation of such beams.
\par
\begin{figure}[htbp!]
\centering
\includegraphics[width=1\columnwidth, keepaspectratio]{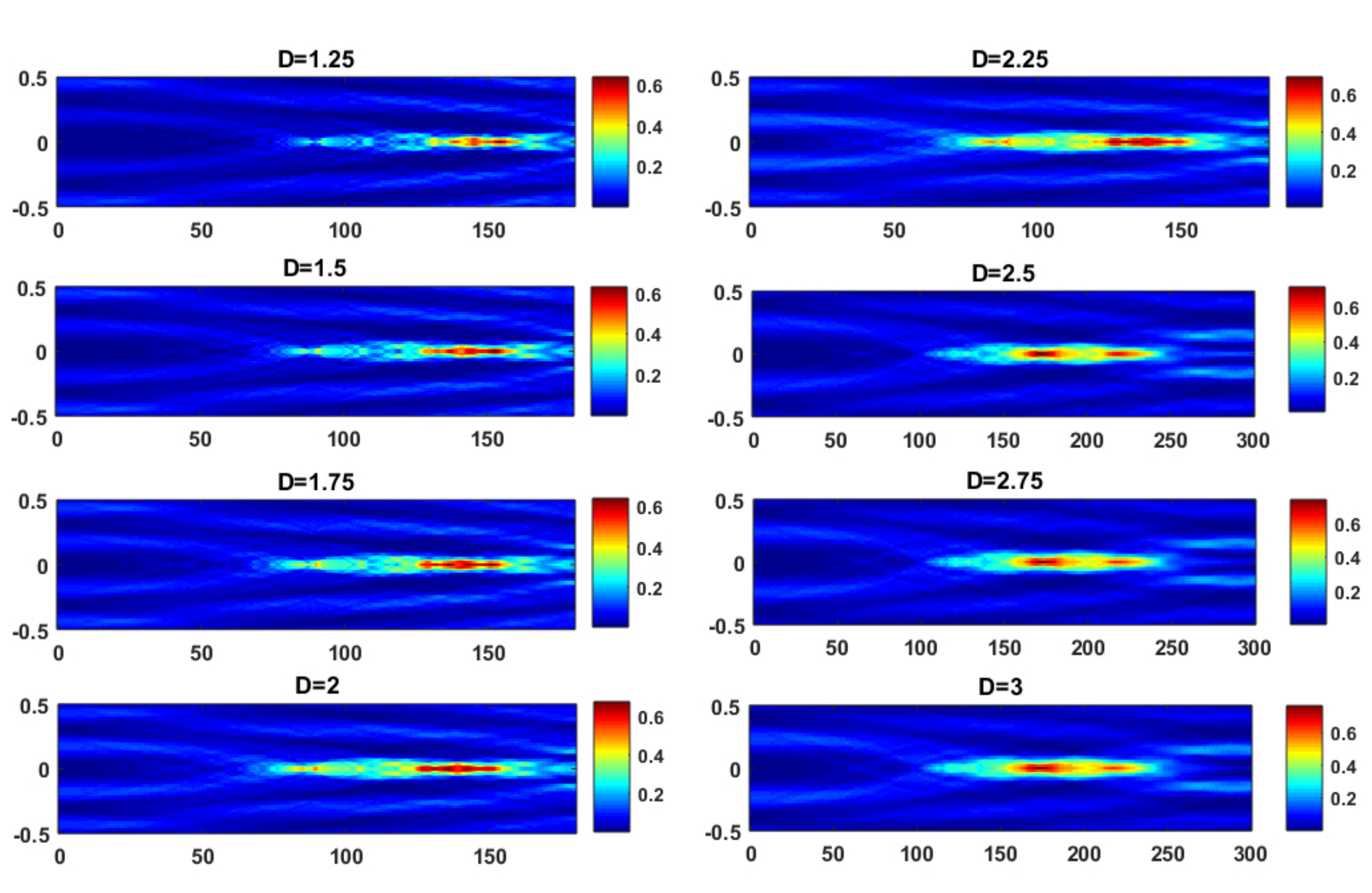}

\caption{XZ intensity profiles of s-SFBB for different dimensions}
\label{fig:XZ SFBB}
\end{figure}

Furthermore, this beam can travel easily at longer distances without changing its intensity and phase profiles and can have any arbitrary value of the orbital angular momentum. This can be seen in Fig. \ref{fig:XZ SFBB} where we have shown the intensity profiles of the SFBB in the propagation direction for different dimensions. The distance and quality of the signal can be adjusted by carefully controlling the dimensions of the SFBB. Another interesting aspect of this beam is the non-existence of the circular symmetry in case of fractional dimensions. This can lead to some fascinating applications such as the spin-jets for separating two or more particles and refractive index sensors for liquids etc. 

\subsection{Optical Microscopy}
In the design of metalens for optical microscopy, SFBBs can have varying depth of focus depending upon their spatial dimensions. This is evident from Fig. \ref{fig:XZ SFBB} where it is shown that for different values of $\alpha_1, \alpha_2$ and $\alpha_3$ we can have variable focal depths. Thus by carefully controlling these values we can achieve desired magnification without affecting the quality of the image. This phenomenon is quite useful in the design of zoom lens.  
\subsection{Laser Fabrication}
SFBBs can also be used to design lens with variable feature sizes depending upon the dimensions of the beam. Thus, we can achieve variable feature sizes with a single standalone device which will not only reduce the complexity of this process but will also be helpful in the fabrication of more complicated devices in a much shorter time.

\section{Conclusion}
Scalar and vector wave solutions of the cylindrical free space scalar wave equation have been analyzed using the fractional-dimensional approach. Our results show that the proposed SFBB possesses continuous order orbital angular momentum depending upon its spatial dimensions. Furthermore, it was observed that the circular symmetry in the transverse or radial direction breaks in case of fractional dimensions. This aspect combined with the self reconstruction and diffraction-free characteristics, can be exploited in various interesting applications such as optical tweezers for particle micromanipulation and trapping, spin-jets for particle separation and in optical communications.
\par
This work also serves as a call for experiments to generate the proposed solution of space-fractional Bessel beams by making use of a digital micromirror device (DMD) or a spatial light modulator as discussed in \cite{Tao3,Vega1,SLM1,SLM2,SLM3}. The same beam can also be generated using strategies mentioned in \cite{highlyEfficientBessel,HighefficiencyBesselbeamarraygenerationbyHuygensmetasurfaces,wangbessel1,wangbessel2} in order to have on-chip solutions that can be easily employed in a wide variety of applications in different fields of optical engineering like in laser cutting, optical tweezing and optical communications etc.

\section{Acknowledgements}
MZ and MQM are supported by ITU Startup Grant. YSA is supported by SUTD Startup Research Grant (SRT3CI121163).

\appendix

\section{\label{overview}Brief overview of space-fractional Bessel beams in the context of fractional dimensional space}
Fractional-dimensional space approach has attracted widespread attention due to its fundamental importance in the electromagnetic modeling of complex disordered structures. This approach is also useful in modeling the behavior of electromagnetic waves in fractal media. The theoretical basis of this approach is mentioned in \cite{Stillinger, ZubairBook, exactCylindrical}. Each coordinate in the integer dimensional space $(x, y$ and $z)$ is represented by a corresponding coordinate in fractional space $(\alpha_1,\alpha_2,$ and $\alpha_3)$. These coordinates can have any fractional value between $0$ and $1$ $(0\leq\alpha_1,\alpha_2,\alpha_3\leq1)$ such that the total dimension of space $D=\alpha_1+\alpha_2+\alpha_3$. For normal three-dimensional space $\alpha_1=\alpha_2=\alpha_3=1$. In this work we have modeled the behaviour of Bessel beams in fractional dimensional space by deriving the exact analytical solution of the fractional Helmholtz equation (defined using the space-fractional cylindrical del operator) termed here as the space-fractional Bessel beam (SFBB). This solution satisfies the fractional Helmholtz equation for all spatial dimensions $(0\leq D\leq3)$. Furthermore, it reduces to ordinary integer Bessel beam for $D=3$. Hence, this solution can be considered as a generalized solution of the fractional Helmholtz equation for any arbitrary spatial dimension. Detailed derivation of this solution and the space-fractional cylindrical del and Laplacian operators is mentioned in the subsequent sections of the Appendix.       

\section{\label{DelOperator}Cylindrical del operator for fractional dimensional space}
Cylindrical coordinates $(\rho,\phi,z)$ can be represented using the Cartesian counterparts $(x,y,z)$ as
\begin{equation}\label{eq:A1}
    \rho=\sqrt{x^2+y^2},\: \phi=\tan^{-1}\frac{y}{x}\: and\: z=z.
\end{equation}
Likewise, we can convert the differentials from Cylindrical to Cartesian coordinate system using the following transformation matrix,
\begin{equation}\label{eq:A6}
    \begin{bmatrix}
    \frac{\partial}{\partial x} \\ 
    \frac{\partial}{\partial y} \\
    \frac{\partial}{\partial z}
    \end{bmatrix}=\begin{bmatrix}
    \cos(\phi) & -\frac{\sin(\phi)}{\rho} & 0\\
    \sin(\phi) & \frac{\cos(\phi)}{\rho} & 0\\
    0 & 0 & 1
    \end{bmatrix}\begin{bmatrix}
    \frac{\partial}{\partial\rho}\\
    \frac{\partial}{\partial\phi}\\
    \frac{\partial}{\partial z}
    \end{bmatrix}
\end{equation}
This implies that,
\begin{equation}\label{eq:A7}
\frac{\partial}{\partial x}=\cos\phi\frac{\partial}{\partial\rho}-\frac{\sin\phi}{\rho}\frac{\partial}{\partial\phi}    
\end{equation}
and,
\begin{equation}\label{eq:A8}
    \frac{\partial}{\partial y}=\sin\phi\frac{\partial}{\partial\rho}+\frac{\cos\phi}{\rho}\frac{\partial}{\partial\phi}
\end{equation}
Del operator in Cartesian coordinates for fractional dimensional space is given by,
\begin{equation}\label{eq:A9}
    \begin{split}
        \boldsymbol{\nabla}={}& \left(\frac{\partial}{\partial x}+\frac{1}{2}\frac{\alpha_1-1}{x}\right)\boldsymbol{x}+\left(\frac{\partial}{\partial y}+\frac{1}{2}\frac{\alpha_2-1}{y}\right)\boldsymbol{y}+\\&\left(\frac{\partial}{\partial z}+\frac{1}{2}\frac{\alpha_3-1}{z}\right)\boldsymbol{z}
    \end{split}
\end{equation}
Here, we can see that,
\begin{equation}\label{eq:A10}
    A_x=\frac{\partial}{\partial x}+\frac{1}{2}\frac{\alpha_1-1}{x}
\end{equation}
and,
\begin{equation}\label{eq:A11}
    A_y=\frac{\partial}{\partial y}+\frac{1}{2}\frac{\alpha_2-1}{y}
\end{equation}
After substituting the known values, we finally get,
\begin{equation}\label{eq:A12}
    A_x=\cos\phi\frac{\partial}{\partial\rho}-\frac{\sin\phi}{\rho}\frac{\partial}{\partial\phi}+\frac{1}{2}\frac{\alpha_1-1}{\rho \cos\phi}
\end{equation}
and,
\begin{equation}\label{eq:A13}
    A_y=\sin\phi\frac{\partial}{\partial\rho}+\frac{\cos\phi}{\rho}\frac{\partial}{\partial\phi}+\frac{1}{2}\frac{\alpha_2-1}{\rho \sin\phi}
\end{equation}
The transformation matrix for converting Cartesian coordinates to cylindrical coordinates is given below,
\begin{equation}\label{eq:A14}
    \begin{bmatrix}
    A_\rho\\
    A_\phi\\
    A_z
    \end{bmatrix}=\begin{bmatrix}
    \cos\phi & \sin\phi & 0\\
    -\sin\phi & \cos\phi & 0\\
    0 & 0 & 1
    \end{bmatrix}\begin{bmatrix}
    A_x \\
    A_y\\
    A_z
    \end{bmatrix}
\end{equation}
By making use of this matrix, we finally get,
\begin{equation}\label{eq:A15}
    A_\rho=\frac{\partial}{\partial\rho}+\frac{1}{2\rho}(\alpha_1+\alpha_2-2),
\end{equation}
\begin{equation}\label{eq:A16}
    A_\phi=\frac{1}{\rho}\frac{\partial}{\partial\phi}-\frac{1}{2\rho}\left((\alpha_1-1)\tan\phi-(\alpha_2-1)\cot\phi\right)
\end{equation}
and,
\begin{equation}\label{eq:A17}
    A_z=\frac{\partial}{\partial z}+\frac{1}{2}\frac{\alpha_3-1}{z}
\end{equation}
Hence, we can now represent the Cylindrical Del operator sing the fractional dimensional approach as follows,
\begin{equation}\label{eq:A18}
    \begin{split}
        \nabla={}&\left(\frac{\partial}{\partial\rho}+\frac{1}{2\rho}(\alpha_1+\alpha_2-2)\right)\boldsymbol{\rho}+\\&
        \left(\frac{1}{\rho}\frac{\partial}{\partial\phi}-\frac{1}{2\rho}\left((\alpha_1-1)\tan\phi-(\alpha_2-1)\cot\phi\right)\right)\boldsymbol{\phi}+\\&
        \left(\frac{\partial}{\partial z}+\frac{1}{2}\frac{\alpha_3-1}{z}\right)\boldsymbol{z}
    \end{split}
\end{equation}

\section{Laplacian operator in cylindrical coordinates in fractional dimensional space}
Transformation matrix for Cartesian to Cylindrical coordinates can be written as
\begin{equation}\label{eq:B1}
    \begin{bmatrix}
\frac{\partial}{\partial x} \\ \frac{\partial}{\partial y}
\end{bmatrix}=\begin{bmatrix}
\cos(\phi) & -\frac{\sin(\phi)}{\rho} \\
\sin(\phi) & \frac{\cos(\phi)}{\rho}
\end{bmatrix}\begin{bmatrix}
\frac{\partial}{\partial\rho} \\
\frac{\partial}{\partial\phi}
\end{bmatrix}
\end{equation}
This implies that,
\begin{equation}\label{eq:B2}
\frac{\partial}{\partial x}=\cos(\phi)\frac{\partial}{\partial\rho}-\frac{\sin(\phi)}{\rho}\frac{\partial}{\partial\phi}    
\end{equation}
and,
\begin{equation}\label{eq:B3}
    \frac{\partial}{\partial y}=\sin(\phi)\frac{\partial}{\partial\rho}+\frac{\cos(\phi)}{\rho}\frac{\partial}{\partial\phi}
\end{equation}
Laplacian operator can be represented using the fractional dimensional approach in terms of the Cartesian coordinate system as shown below, 
\begin{equation}\label{eq:B4}
\begin{split}
        \nabla^2={}&\frac{\partial^2}{\partial x^2}+\frac{\alpha_1-1}{x}\frac{\partial}{\partial x}+\frac{\partial^2}{\partial y^2}+\frac{\alpha_2-1}{y}\frac{\partial}{\partial y}+\frac{\partial^2}{\partial z^2}+\\&\frac{\alpha_3-1}{z}\frac{\partial}{\partial z}
\end{split}
\end{equation}
Multiplying Eq. \ref{eq:B2} with itself, we get,
\begin{equation}\label{eq:B5}
\begin{split}
    \frac{\partial ^2}{\partial x^2}={}&\cos^2(\phi)\frac{\partial^2}{\partial \rho^2}-\frac{2\sin(\phi)\cos(\phi)}{\rho}\frac{\partial^2}{\partial\rho\partial\phi}+\\&\frac{2\sin(\phi)\cos(\phi)}{\rho^2}\frac{\partial}{\partial\phi}+\frac{\sin^2(\phi)}{\rho}\frac{\partial}{\partial\rho}+\\&\frac{\sin^2(\phi)}{\rho^2}\frac{\partial^2}{\partial \phi^2}  
\end{split}
\end{equation}
Similarly,
\begin{equation}\label{eq:B6}
    \begin{split}
        \frac{\partial^2}{\partial y^2}={}&\sin^2(\phi)\frac{\partial^2}{\partial \rho^2}+\frac{2\sin(\phi)\cos(\phi)}{\rho}\frac{\partial^2}{\partial\rho\partial\phi}-\\&\frac{2\sin(\phi)\cos(\phi)}{\rho^2}\frac{\partial}{\partial\phi}+\frac{\cos^2(\phi)}{\rho}\frac{\partial}{\partial\rho}+\\&\frac{\cos^2(\phi)}{\rho^2}\frac{\partial^2}{\partial \phi^2}
    \end{split}
\end{equation}
After substituting the values of these differentials in Eq. \ref{eq:B4}, we get,
\begin{equation}\label{eq:B7}
    \begin{split}
        \nabla^2={}&\frac{\partial^2}{\partial \rho^2}+\frac{1}{\rho}\frac{\partial}{\partial\rho}+\frac{1}{\rho^2}\frac{\partial^2}{\partial \phi^2}+\frac{\alpha_1-1}{\rho \cos(\phi)}\\&\bigg(\cos(\phi)\frac{\partial}{\partial\rho}-\frac{\sin(\phi)}{\rho}\frac{\partial}{\partial\phi}\bigg)+\frac{\alpha_2-1}{\rho \sin(\phi)}\\&\bigg(\sin(\phi)\frac{\partial}{\partial\rho}+\frac{\cos(\phi)}{\rho}\frac{\partial}{\partial \phi}\bigg)+\frac{\partial^2}{\partial z^2}+\frac{\alpha_3-1}{z}\frac{\partial}{\partial z}
    \end{split}
\end{equation}
After further simplification we get the following expression for Laplacian operator in cylindrical coordinates in fractional dimensional space,
\begin{equation}\label{eq:B8}
    \begin{split}
        \nabla^2={}&\frac{\partial^2}{\partial \rho^2}+\frac{1}{\rho}(\alpha_1+\alpha_2-1)\frac{\partial}{\partial\rho}+\frac{1}{\rho^2}\bigg[\frac{\partial^2}{\partial \phi^2}-\\&\{(\alpha_1-1)\tan(\phi)-(\alpha_2-1)\cot(\phi)\}\bigg]\frac{\partial}{\partial \phi}+\\&\frac{\partial^2}{\partial z^2}+\frac{\alpha_3-1}{z}\frac{\partial}{\partial z}
    \end{split}
\end{equation}

\section{Solution of free space scalar wave equation in terms of fractional dimensional space}
Free space scalar wave equation for lossless media in the absence of source can be written as
\begin{equation}\label{eq:C1}
    \nabla^2 \textbf{E} + k^2 \textbf{E} = 0
\end{equation}
\begin{equation}\label{eq:C2}
    \nabla^2 \textbf{H} + k^2 \textbf{H} = 0
\end{equation}
$\textbf{E}$ and $\textbf{H}$ in the above equation are used to represent the complex electric and magnetic field, $k =\frac{2\pi}{\lambda}$ is the free space wave vector. Time dependency $ e^{j\omega t} $ is suppressed in this discussion. $\nabla ^2 $ or Laplacian operator used in these equations is defined in rectangular coordinate system using the following expression
\begin{equation}\label{eq:C3}
\begin{split}
    \nabla_D^2 ={}& \frac{\partial^2}{\partial x^2} \space +  \frac{\alpha_1-1}{x}\frac{\partial}{\partial x}+ (\frac{\partial^2}{\partial y^2}+\frac{\alpha_2-1}{y}\frac{\partial}{\partial y})+     \\&\frac{\partial^2}{\partial z^2}+\frac{\alpha_3-1}{z}\frac{\partial}{\partial z}
\end{split}
\end{equation}
For cylindrical coordinate system this Laplacian operator takes the form \cite{SFBB,ZubairBook}
\begin{equation}\label{eq:C4}
\begin{split}
    \nabla_D^2 ={}& \frac{\partial^2}{\partial \rho^2} \space +  \frac{1}{\rho}(\alpha_1+\alpha_2-1)\frac{\partial}{\partial \rho} \quad+ \frac{1}{\rho^2}\bigg(\frac{\partial^2}{\partial \phi^2}-\\&\{(\alpha_1-1)\tan\phi - (\alpha_2-1)\cot\phi\}\frac{\partial}{\partial\phi}\bigg) +     \\& \frac{\partial^2}{\partial z^2}+\frac{\alpha_3-1}{z}\frac{\partial}{\partial z}
\end{split}
\end{equation}
$\rho$,$\phi$ and $ z $ in \ref{eq:C4} represent each coordinate of the cylindrical coordinate system. Other parameters that are used the define the spatial dimensions are $0\leq\alpha_1\leq1$,$0\leq\alpha_2\leq1$,$0\leq\alpha_3\leq1$. The overall dimension of the beam is $D=\alpha_1+\alpha_2+\alpha_3$ where each one of them acts independently. In order to find the exact solution that satisfies the free space scalar wave equation for fractional dimensional space, we need to start with either Eq. \ref{eq:C1} or Eq. \ref{eq:C2} and replace the complex fields at the end to get the solution of the remaining field due to duality. We will start with \textbf{E} and will replace it with \textbf{H} once its solution is derived.
Electric Field in cylindrical coordinate system is expressed as
\begin{equation}\label{eq:C5}
\begin{split}
        \textbf{E}(\rho,\phi,z)={}&\hat{a}_\rho E_\rho(\rho,\phi,z)+\hat{a}_\phi E_\phi(\rho,\phi,z)+\\&\hat{a}_z E_z(\rho,\phi,z)
\end{split}
\end{equation}
After substituting the expression for $ \textbf{E}(\rho,\phi,z)$ in Eq. \ref{eq:C1}, we will get,
\begin{equation}\label{eq:C6}
\begin{split}
    \nabla_D^2 (\hat{a}_\rho E_\rho+{}&\hat{a}_\phi E_\phi+\hat{a}_z E_z) + \\&k^2 (\hat{a}_\rho E_\rho+\hat{a}_\phi E_\phi+\hat{a}_z E_z) = 0
\end{split}
\end{equation}
It should be noted that
\begin{equation}\label{eq:C7}
    \nabla_D^2(\hat{a}_\rho E_\rho)\neq\hat{a}_\rho\nabla_D^2 E_\rho
\end{equation}
\begin{equation}\label{eq:C8}
    \nabla_D^2(\hat{a}_\phi E_\phi)\neq\hat{a}_\phi\nabla_D^2 E_\phi
\end{equation}
\begin{equation}\label{eq:C9}
    \nabla_D^2(\hat{a}_z E_z)=\hat{a}_z\nabla_D^2 E_z
\end{equation}
therefore, we cannot reduce Eq. \ref{eq:C6} in terms of simple scalar expressions, however, it is possible to achieve partial coupled scalar differential equations.For the ease of calculations, only those solutions are considered that comply with the following free space scalar wave equation:
\begin{equation}\label{eq:C10}
    \nabla_D^2 \psi(\rho,\phi,z) + k^2 \psi(\rho,\phi,z) = 0
\end{equation}
$\psi(\rho,\phi,z)$ in the above equation is used to represent the scalar potential such as a scalar field and can be easily converted to the vector counterpart by the help of Hertz potentials and Lorenz gauge. Upon expansion of Eq. \ref{eq:C10} following expression is achieved,
\begin{equation}\label{eq:C11}
    \begin{split}
    {}&\frac{\partial^2 \psi}{\partial \rho^2} \space +  \frac{1}{\rho}(\alpha_1+\alpha_2-1)\frac{\partial \psi}{\partial \rho} \quad+ \\&\frac{1}{\rho^2}(\frac{\partial^2 \psi}{\partial \phi^2}-\{(\alpha_1-1)\tan\phi - (\alpha_2-1)\cot\phi\}\frac{\partial \psi}{\partial\phi}) \quad+     \\& \frac{\partial^2 \psi}{\partial z^2}+\frac{\alpha_3-1}{z}\frac{\partial \psi}{\partial z} + k^2\psi=0
\end{split}
\end{equation}
Now, by the help of method of separation of variables, we can define,
\begin{equation}\label{eq:C12}
    \psi(\rho,\phi,z)=f(\rho)g(\phi)h(z)
\end{equation}
this results in three ordinary differential equations which are mentioned below
\begin{equation}\label{eq:C13}
    \bigg[\rho^2\frac{d^2}{d \rho^2} \space +  {\rho}(\alpha_1+\alpha_2-1)\frac{d}{d \rho}+(k_\rho\rho)^2-m^2\bigg]f(\rho)=0
\end{equation}
\begin{equation}\label{eq:C14}
    \bigg[\frac{d^2}{d \phi^2}-\{(\alpha_1-1)\tan\phi - (\alpha_2-1)\cot\phi\}\frac{d}{d\phi}+m^2\bigg]g(\phi)=0
\end{equation}
\begin{equation}\label{eq:C15}
    \bigg[\frac{d^2 }{d z^2}+\frac{\alpha_3-1}{z}\frac{d}{dz} + k^2_z\bigg]h(z)=0
\end{equation}
where, 
\begin{equation}\label{eq:C16}
    k^2_\rho+k^2_z=k^2
\end{equation}

Eq. \ref{eq:C13} can be solved for $f(\rho)$.It can be written as
\begin{equation}\label{eq:C17}
     \bigg[ \rho^2\frac{d^2}{d\rho^2}+\alpha_\rho\frac{d}{d\rho}+(\beta\rho^l+c)\bigg]f(\rho)=0
\end{equation}
where, $\alpha_\rho=\alpha_1+\alpha_2-1$, $\beta=k^2_\rho$, $c=-m^2$ and $l=2$.
This expression is similar to the ordinary Bessel equation and it has solution of the form
\begin{equation}\label{eq:C18}
     \rho^{1-\frac{a}{2}}\left[C_1 J_{v_1} \left(\frac{2}{l}\sqrt{b}\rho^\frac{l}{2}\right)+C_2 Y_{v_1} \left(\frac{2}{l}\sqrt{b}\rho^\frac{l}{2}\right)\right]
\end{equation}
where, $v_1=\frac{1}{l}\sqrt{\left(1-a\right)^2-4c}$.
\newline Hence, the final solution of Eq. \ref{eq:C13} becomes,
\begin{equation}\label{eq:C19}
    f_1\left(\rho\right)=\rho^{1-\frac{\left(\alpha_1+\alpha_2\right)}{2}}\left[C_1 J_{v1} \left(k_\rho\rho\right)+C_2 Y_{v1} \left(k_\rho\rho\right)\right]
\end{equation}
or,
\begin{equation}\label{eq:C20}
    f_2\left(\rho\right)=\rho^{1-\frac{\left(\alpha_1+\alpha_2\right)}{2}}\left[D_1 H_v^{\left(1\right)} \left(k_\rho\rho\right)+D_2 H_v^{\left(2\right)} \left(k_\rho\rho\right)\right]
\end{equation}
where, $v_1=\frac{1}{2}\sqrt{\left(2-\alpha_1-\alpha_2\right)^2+4m^2}$. $J_{v_1}\left(k_\rho\rho\right)$ and $Y_{v_1}\left(k_\rho\rho\right)$ in Eq. \ref{eq:C19}are Bessel functions of first and second kind of the order $v$ respectively. $H_v^{\left(1\right)}{\left(k_\rho\rho\right)}$ and $H_v^{\left(2\right)}{\left(k_\rho\rho\right)}$ in Eq. \ref{eq:C20}are known as Hankel functions.They generally replace Bessel beams in case of travelling waves.
\newline Now, we have to find the solution for $g\left(\phi\right)$ in Eq. \ref{eq:C14}. This equation resembles the equation mentioned below
\begin{equation}\label{eq:C21}
\begin{split}
        y_{xx}''+(a \tan x+{}&b \cot x ) y_x'+\\&(\alpha \tan^2x+\beta \cot^2x+\gamma) y=0
\end{split}
\end{equation}
where, $y=g$, $x=\phi$, $a=1-\alpha_1$, $b=\alpha_2-1$, $\alpha=0$, $\beta=0$ and $\gamma=m^2$. Using the substitution $\zeta=\sin^2x$ and $y=w\sin^px\cos^qx$ where $p$ and $q$ can be obtained by solving the following simultaneous linear equations 
\newline    \begin{center}$p^2+(b-1)p+\beta=0, q^2-(a+1)q+\alpha=0$\end{center}
This leads to the equation
\begin{equation}\label{eq:C22}
\begin{split}
        {}&4\zeta\left(\zeta-1\right)w_{\zeta\zeta}''+2\big[\left(2p+2q+2+b-a\right)\zeta-2p-\\&b-1\big]w_\zeta'+\left(2pq+p+q+bq-ap-\gamma\right)w=0
\end{split}
\end{equation}
where, $\zeta=\sin^2(\phi)$, Eq. \ref{eq:C22} is similar to hypergeometric equation given below
\begin{equation}\label{eq:C23}
   x\left(x-1\right)y_{xx}''+\left[\left(\alpha+\beta+1\right)x-\gamma\right]y_x'+\alpha\beta y=0 
\end{equation}
where, $x=\zeta$, $y=w$, $\alpha+\beta+1=\frac{1}{2}\left(8-\alpha_2-\alpha_1\right)$, $\gamma=\frac{1}{2}\left(4-\alpha_2\right)$ and $\alpha\beta=\frac{1}{4}\left(8-2\alpha_1-2\alpha_2-m^2\right)$. So, the final expression for $w$ can be written as
\begin{equation}\label{eq:C24}
\begin{split}
        w={}&C_1F\left(\alpha,\beta,\gamma;\zeta\right)+\\&C_2\zeta^{1-\gamma}F\left(\alpha-\gamma+1,\beta-\gamma+1,2-\gamma;\zeta\right)
\end{split}
\end{equation}
Hence, the final solution of Eq. \ref{eq:C14} becomes 
\begin{equation}\label{eq:C25}
\begin{split}
   g(\phi)={}& \sin^{2-\alpha_2}\phi \cos^{2-\alpha_1}\phi \times \big(C_3F\left(\alpha,\beta,\gamma;\zeta\right)+\\& C_4\zeta^{1-\gamma}F\left(\alpha-\gamma+1,\beta-\gamma+1,2-\gamma;\zeta\right)\big)
\end{split}
\end{equation}
Finally, solving for $h(z)$ in Eq. \ref{eq:C15}. This equation can be reduced to Hypergeometric equation given below,
\begin{equation}\label{eq:C26}
    (a_2x+b_2)y_{xx}''+(a_1x+b_1)y_{x}'+(a_0x+b_0)y=0
\end{equation}
where, $a_2=1 ,a_1=0, a_0=k_z^2, b_2=0, b_1=\alpha_3-1, b_0=0, y=h(z) and x=z$. Solution to this equation is of the form
\begin{equation}\label{eq:C27}
    y=e^{kz}U(a,b;v_2)
\end{equation}
where, $k=\frac{\sqrt{D}-a_1}{2a_2}, D=a_1^2-4a_0a_2, v_2=\frac{z-\mu}{\lambda}, \mu=\frac{-b_2}{a_2}, \lambda=\frac{-a_2}{2a_2k+a_1},   a=\frac{B(k)}{2a_2k+a_1}, b=\frac{a_2b_1-a_1b_2}{a_2^2}, B(k)=b_2k^2+b_1k+b_0$ and $U(a,b,v_2)$ is the kummer Hypergeometric function. Hence,
\begin{equation}\label{eq:C28}
    h\left(z\right)=e^{kz}U(a,b,v_2)
\end{equation}
So, the final solution for free space Cylindrical Scalar Wave Equation using the fractional dimensional approach can be written as,
\begin{equation}\label{eq:C29}
\begin{split}
     \psi(\rho,\phi,z)={}& \rho^{1-\frac{\left(\alpha_1+\alpha_2\right)}{2}}\left[C_1 J_{v_1} \left(k_\rho\rho\right)+C_2 Y_{v_1} \left(k_\rho\rho\right)\right] \times\\& \sin^{2-\alpha_2}\phi \cos^{2-\alpha_1}\phi \big[C_3F\left(\alpha,\beta,\gamma;\zeta\right)+\\& C_4\zeta^{1-\gamma}F\left(\alpha-\gamma+1,\beta-\gamma+1,2-\gamma;\zeta \right)\big] \\&  \times e^{kz}U(a,b,v_2)
\end{split}   
\end{equation}
\section{\label{DetailedVectorAnalysis}Vector wave analysis of the space-fractional Bessel beam}
Scalar space-fractional Bessel beam can be transformed into vector representation by making use of the Hertz vector potentials and the Lorenz Gauge condition as follows,
\begin{equation}\label{eq:D10}
    \boldsymbol{E_e}=\boldsymbol{\nabla_D}(\boldsymbol{\nabla_D}.\boldsymbol{\Pi_e})-k^2\boldsymbol{\Pi_e}
\end{equation}
\begin{equation}\label{eq:D11}
    \boldsymbol{H_e}=-ik\frac{1}{\eta}(\boldsymbol{\nabla_D}\times\boldsymbol{\Pi_e})
\end{equation}
\begin{equation}\label{eq:D12}
    \boldsymbol{E_m}=ik\eta(\boldsymbol{\nabla_D}\times\boldsymbol{\Pi_m})
\end{equation}
\begin{equation}\label{eq:D13}
    \boldsymbol{H_m}=\boldsymbol{\nabla_D}(\boldsymbol{\nabla_D}.\boldsymbol{\Pi_m})-k^2\boldsymbol{\Pi_m}
\end{equation}
where,  $e$ and $m$ in the subscript are used to denote the TMn and TEn modes respectively and the total field can be calculated by adding both of these modes, $k=\frac{2\pi}{\lambda}$ is the wave vector and, $\boldsymbol{\Pi_e}$ and $\boldsymbol{\Pi}_m$ are the corresponding Hertz vector potentials which can be expressed using the equation mentioned below,
\begin{widetext}
\begin{equation}\label{eq:D14}
\begin{split}
     \boldsymbol{\Pi_{e/m}}={}&\Pi_{e/m} \boldsymbol{\hat{z}}=A_{e/m}\rho^{1-\frac{(\alpha_1+\alpha_2)}{2}}\bigg[C_1J_{v_1}(k_\rho\rho)\bigg]\times \sin^{2-\alpha_2}\phi \cos^{2-\alpha_1}\phi \times\\&
     \bigg[C_3F(\alpha,\beta,\gamma;\zeta)+C_4\zeta^{1-\gamma}F(\alpha-\gamma+1,\beta-\gamma+1,2-\gamma;\zeta)\bigg]\times \\&e^{kz}U(a,b;v_2)\boldsymbol{\hat{z}}
\end{split}
\end{equation}
\end{widetext}
Here, $\Pi_{e/m}$ is the scalar electric/magnetic potential and it can be expressed in a similar way as the s-SFBB. Hankel functions and time dependency is suppressed in this case for simplicity.
\newline Del operator in cylindrical coordinates for fractional dimensional space is given by (see Appendix \ref{DelOperator}),
\begin{widetext}
\begin{equation}\label{eq:D16}
\begin{split}
        \boldsymbol{\nabla_D}={}&\bigg[\frac{\partial}{\partial \rho}+\frac{(\alpha_1+\alpha_2-2)}{2\rho}\bigg]\boldsymbol{\hat{\rho}}+\bigg[\frac{1}{\rho}\frac{\partial}{\partial\phi}-\frac{1}{2\rho}\{(\alpha_1-1)\tan\phi-(\alpha_2-1)\cot(\phi)\}\bigg]\boldsymbol{\hat{\phi}}+\\&
        \bigg[\frac{\partial}{\partial z}+\frac{1}{2}\frac{(\alpha_3-1)}{z}\bigg]\boldsymbol{\hat{z}}
\end{split}
\end{equation}
\end{widetext}
Finding the dot product of the electric and magnetic scalar potential function, we get,
\begin{widetext}
\begin{equation}\label{eq:D17}
    \begin{split}
        \boldsymbol{\nabla.\Pi_{e/m}}={}&\left(\frac{\partial}{\partial z}+\frac{\alpha_3-1}{z}\right)\Pi_{e/m}\boldsymbol{\hat{z}}
    \end{split}
\end{equation}

\begin{equation}\label{eq:D18}
    \begin{split}
        \boldsymbol{\nabla.\Pi_{e/m}}={}&\bigg[A_{e/m}\rho^{1-\frac{(\alpha_1+\alpha_2)}{2}}\bigg(C_1J_{v_1}(\beta_\rho\rho)\bigg)\times \sin^{2-\alpha_2}\phi \cos^{2-\alpha_1}\phi\times
        \bigg(C_3F(\alpha,\beta,\gamma;\zeta)+\\&C_4\zeta^{1-\gamma}F(\alpha-\gamma+1,\beta-\gamma+1,2-\gamma;\zeta)\bigg)\times\bigg(ke^{kz}U(a,b,v_2)-\\&ae^{kz}U(a+1,b+1,v_2)\bigg)\bigg]+ \bigg[\frac{(\alpha_3-1)}{2}A_{e/m}\rho^{1-\frac{(\alpha_1+\alpha_2)}{2}}(C_1J_{v_1}(\beta_\rho\rho))\times\\& \sin^{2-\alpha_2}\phi \cos^{2-\alpha_1}\phi\times
        \bigg(C_3F(\alpha,\beta,\gamma;\zeta)+\\&C_4\zeta^{1-\gamma}F(\alpha-\gamma+1,\beta-\gamma+1,2-\gamma,\zeta)\bigg)\times
        (z^{-1}e^{kz}U(a,b,v_2))\bigg]
    \end{split}
\end{equation}
\end{widetext}

After calculating the dot products, we can now determine the analytical expressions for the $\rho$, $\phi$ and $z$ components of the TMn mode of the electric and magnetic fields respectively. This can be done by the help of Eq. \ref{eq:D10} and Eq. \ref{eq:D11}, as follows,
\begin{widetext}
\begin{equation}\label{eq:D21}
    \begin{split}
        E_{\rho e}{}&=[\boldsymbol{\nabla_D(\nabla_D.\Pi_e)}]_\rho=\bigg[\frac{\partial}{\partial\rho}+\frac{1}{2\rho}(\alpha_1+\alpha_2-2)\bigg](\boldsymbol{\nabla_D.\Pi_e})
    \end{split}
\end{equation}
\begin{equation}\label{eq:D22}
    \begin{split}
        E_{\rho e}={}&\bigg[A_e\bigg(\bigg(1-\frac{(\alpha_1+\alpha_2)}{2}\bigg)\rho^{-\frac{(\alpha_1+\alpha_2)}{2}}C_1\bigg(J_{v_1}(\beta_\rho\rho)\bigg)+\frac{C_1}{2}\rho^{1-\frac{(\alpha_1+\alpha_2)}{2}}\times\\&\beta_\rho(J_{v_1-1}(\beta_\rho\rho)-J_{v_1+1}(\beta_\rho\rho))\bigg)\times\bigg(\sin^{2-\alpha_2}\phi\cos^{2-\alpha_1}\phi\\&\bigg(C_3F(\alpha,\beta,\gamma;\zeta)+C_4\zeta^{1-\gamma}F(\alpha-\gamma+1,\beta-\gamma+1,2-\gamma;\zeta)\bigg)\bigg)\times\\&\bigg(ke^{kz}U(a,b;v_2)-ae^{kz}U(a+1,b+1;v_2)\bigg)\bigg]+ \bigg[\frac{(\alpha_3-1)}{2}A_e\\&\bigg(\bigg(1-\frac{(\alpha_1+\alpha_2)}{2}\bigg)\rho^{-\frac{(\alpha_1+\alpha_2)}{2}}C_1\bigg(J_{v_1}(\beta_\rho\rho)\bigg)+\frac{C_1}{2}\rho^{1-\frac{(\alpha_1+\alpha_2)}{2}}\beta_\rho\\&\bigg(J_{v_1-1}(\beta_\rho\rho)-J_{v_1+1}(\beta_\rho\rho)\bigg)\bigg)\times\bigg(\sin^{2-\alpha_2}\phi\cos^{2-\alpha_1}\phi\times\bigg(C_3\\&F(\alpha,\beta,\gamma;\zeta)+C_4\zeta^{1-\gamma}F(\alpha-\gamma+1,\beta-\gamma+1,2-\gamma;\zeta)\bigg)\bigg)\times\\&\bigg(z^{-1}e^{kz}U(a,b;v_2)\bigg)\bigg]+
        \bigg[\frac{1}{2\rho}(\alpha_1+\alpha_2-2)(\boldsymbol{\nabla_D.\Pi_e})\bigg]
    \end{split}
\end{equation}

\begin{equation}\label{eq:D23}
    \begin{split}
        E_{\phi e}{}&=[\boldsymbol{\nabla_D(\nabla_D.\Pi_e)}]_{\phi}\\&=\bigg[\frac{1}{\rho}\frac{\partial}{\partial\phi}-\frac{1}{2\rho}[(\alpha_1-1)\tan\phi-(\alpha_2-1)\cot\phi]\bigg](\boldsymbol{\nabla_D.\Pi_e})
    \end{split}
\end{equation}

\begin{equation}\label{eq:D24}
    \begin{split}
        E_{\phi e}={}&\bigg[A_e\rho^{-\frac{(\alpha_1+\alpha_2)}{2}}\bigg(C_1J_{v_1}(\beta_\rho\rho)\bigg)\times\bigg(\bigg((2-\alpha_2)\cos^{3-\alpha_1}\phi\sin^{1-\alpha_2}\phi-\\&(2-\alpha_1)\cos^{1-\alpha_1}\phi\sin^{3-\alpha_2}\phi\bigg)C_3F(\alpha,\beta,\gamma;\zeta)+2\sin^{3-\alpha_2}\phi\cos^{3-\alpha_1}\phi\\&\frac{\alpha\beta}{\gamma}C_3F(\alpha+1,\beta+1,\gamma+1;\zeta)+\bigg((4-\alpha_2-2\gamma)\cos^{3-\alpha_1}\phi\sin^{3-\alpha_2-2\gamma}\phi-\\&(2-\alpha_1)\cos^{1-\alpha_1}\phi\sin^{5-\alpha_2-2\gamma}\phi\bigg)C_4
        F(\alpha-\gamma+1,\beta-\gamma+1,2-\gamma;\zeta)+\\&2\sin^{5-\alpha_2-2\gamma}\phi\cos^{3-\alpha_1}\phi
        \frac{(\alpha-\gamma+1)(\beta-\gamma+1)}{(2-\gamma)}C_4\\&F(\alpha-\gamma+2,\beta-\gamma+2,3-\gamma;\zeta)\bigg)\times
        \bigg(ke^{kz}U(a,b;v_2)-\\&ae^{kz}U(a+1,b+1;v_2)\bigg)\bigg]+
        \bigg[\frac{(\alpha_3-1)}{2}A_e\rho^{-\frac{(\alpha_1+\alpha_2)}{2}}\bigg(C_1J_{v_1}(\beta_\rho\rho)\bigg)\times\\&\bigg(\bigg((2-\alpha_2)\cos^{3-\alpha_1}\phi\sin^{1-\alpha_2}\phi-(2-\alpha_1)\cos^{1-\alpha_1}\phi\sin^{3-\alpha_2}\phi\bigg)\\& C_3F(\alpha,\beta,\gamma;\zeta)+2\sin^{3-\alpha_2}\phi\cos^{3-\alpha_1}\phi\times\frac{\alpha\beta}{\gamma}\times \\&C_3F(\alpha+1,\beta+1,\gamma+1;\zeta)+\bigg((4-\alpha_2-2\gamma)\cos^{3-\alpha_1}\phi\sin^{3-\alpha_2-2\gamma}\phi-\\&(2-\alpha_1)\cos^{1-\alpha_1}\phi\sin^{5-\alpha_2-2\gamma}\phi\bigg)C_4F(\alpha-\gamma+1,\beta-\gamma+1,2-\gamma;\zeta)\\&+2\sin^{5-\alpha_2-2\gamma}\phi\cos^{3-\alpha_1}\phi\frac{(\alpha-\gamma+1)(\beta-\gamma+1)}{(2-\gamma)}C_4\\&F(\alpha-\gamma+2,\beta-\gamma+2,3-\gamma;\zeta)\bigg)\times\bigg(z^{-1}e^{kz}U(a,b;v_2)\bigg)\bigg]-\\&\bigg[\frac{1}{2\rho}((\alpha_1-1)\tan\phi-(\alpha_2-1)\cot\phi)(\boldsymbol{\nabla_D.\Pi_e})\bigg]
    \end{split}
\end{equation}
\begin{equation}\label{eq:D25}
    \begin{split}
       E_{ze}{}&=[\boldsymbol{\nabla_D(\nabla_D.\Pi_e)}]_z-k^2\Pi_e=\bigg[\frac{\partial}{\partial z}+\frac{(\alpha_3-1)}{2z}\bigg][\boldsymbol{\nabla_D.\Pi_e}]-k^2\boldsymbol{\Pi_e}
    \end{split}
\end{equation}

\begin{equation}\label{eq:D26}
    \begin{split}
        E_{ze}={}&\bigg[A_e\rho^{1-\frac{(\alpha_1+\alpha_2)}{2}}\bigg(C_1J_{v_1}(\beta_\rho\rho)\bigg)\times\bigg(\sin^{2-\alpha_2}\phi\cos^{2-\alpha_1}\phi\times
        \bigg(C_3F(\alpha,\beta,\gamma;\zeta)+\\&C_4\zeta^{1-\gamma}F(\alpha-\gamma+1,\beta-\gamma+1,2-\gamma;\zeta)\bigg)\times\bigg(k^2e^{kz}U(a,b,v_2)-2ake^{kz}\\&U(a+1,b+1;v_2)+a(a+1)e^{kz}U(a+2,b+2;v_2)\bigg)\bigg]+\\&\bigg[\frac{(\alpha_3-1)}{2}A_e\rho^{1-\frac{(\alpha_1+\alpha_2)}{2}}\bigg(C_1 J_{v_1}(\beta_\rho\rho)\bigg)\times\bigg(C_3F(\alpha,\beta,\gamma;\zeta)+\\&C_4
        \zeta^{1-\gamma}F(\alpha-\gamma+1,\beta-\gamma+1,2-\gamma;\zeta)\bigg)\times\bigg(\frac{e^{kz}(kz-1)}{z^2}U(a,b;v_2)-\\&\frac{ae^{kz}}{z}U(a+1,b+1;v_2)\bigg]+\bigg[\frac{(\alpha_3-1)}{2z}(\boldsymbol{\nabla_D.\Pi_e})\bigg]-
        \bigg[k^2A_e\rho^{1-\frac{(\alpha_1+\alpha_2)}{2}}\\&\bigg(C_1 J_{v_1}(\beta_\rho\rho)\bigg)\times\bigg(\sin^{2-\alpha_2}\phi\cos^{2-\alpha_1}\phi\times\bigg(C_3F(\alpha,\beta,\gamma;\zeta)+\\&C_4\zeta^{1-\gamma}F(\alpha-\gamma+1,\beta-\gamma+1,2-\gamma;\zeta)\bigg)\bigg)\times\bigg(e^{kz}U(a,b;v_2)\bigg)\bigg]
    \end{split}
\end{equation}

\begin{equation}\label{eq:D27}
    \begin{split}
        H_{\rho e}={}&-i\frac{k}{\eta}\bigg[\frac{1}{\rho}\frac{\partial}{\partial \phi}\boldsymbol{\Pi_e}-\frac{\Pi_e}{2\rho}[(\alpha_1-1)\tan\phi-(\alpha_2-1)\cot\phi]\bigg]
    \end{split}
\end{equation}

\begin{equation}\label{eq:D28}
    \begin{split}
        H_{\rho e}={}&\bigg[-i\frac{k}{\eta}A_e\rho^{-\frac{(\alpha_1+\alpha_2)}{2}}\bigg(C_1 J_{v_1}(\beta_\rho\rho)\bigg)\times\bigg(\bigg((2-\alpha_2)\cos^{3-\alpha_1}\phi\sin^{1-\alpha_2}\phi-\\&(2-\alpha_1)\cos^{1-\alpha_1}\phi\sin^{3-\alpha_2}\phi\bigg)C_3F(\alpha,\beta,\gamma;\zeta)+2\sin^{3-\alpha_2}\phi\cos^{3-\alpha_1}\phi\times\\&\frac{\alpha\beta}{\gamma}C_3 F(\alpha+1,\beta+1,\gamma+1;\zeta)+\bigg((4-\alpha_2-2\gamma)\cos^{3-\alpha_1}\phi\sin^{3-\alpha_2-2\gamma}\phi-\\&(2-\alpha_1)\cos^{1-\alpha_1}\phi\sin^{5-\alpha_2-2\gamma}\phi\bigg)C_4 F(\alpha-\gamma+1,\beta-\gamma+1,2-\gamma;\zeta)+2\\&\sin^{5-\alpha_2-2\gamma}\phi\cos^{3-\alpha_1}\phi\frac{(\alpha-\gamma+1)(\beta-\gamma+1)}{(2-\gamma)}C_4 \\&F(\alpha-\gamma+2,\beta-\gamma+2,3-\gamma;\zeta)\bigg)\times\bigg(e^{kz}U(a,b;v_2)\bigg)\bigg]+
        \bigg[i\frac{k}{2\eta}A_e\\&((\alpha_1-1)\tan\phi-(\alpha_2-1)\cot\phi)\rho^{-\frac{(\alpha_1+\alpha_2)}{2}}\bigg(C_1 J_{v_1}(\beta_\rho\rho)\bigg)\times\\&\bigg(\sin^{2-\alpha_2}\phi\cos^{2-\alpha_1}\phi\times\bigg(C_3 F(\alpha,\beta,\gamma;\zeta)+C_4\zeta^{1-\gamma}\\&F(\alpha-\gamma+1,\beta-\gamma+1,2-\gamma;\zeta)\bigg)\times\bigg(e^{kz}U(a,b;v_2)\bigg)\bigg]
    \end{split}
\end{equation}

\begin{equation}\label{eq:D29}
    \begin{split}
        H_{\phi e}={}&i\frac{k}{\eta}\bigg[\frac{\partial}{\partial\rho}\boldsymbol{\Pi_e}+\frac{\boldsymbol{\Pi_e}}{2\rho}(\alpha_1+\alpha_2-2)\bigg]
    \end{split}
\end{equation}

\begin{equation}\label{eq:D30}
    \begin{split}
        H_{\phi e}={}&\bigg[i\frac{k}{\eta}A_e\bigg(\bigg(1-\frac{(\alpha_1+\alpha_2)}{2}\bigg)\rho^{-\frac{(\alpha_1+\alpha_2)}{2}}C_1 J_{v_1}(\beta_\rho\rho)+\frac{C_1}{2}\rho^{1-\frac{(\alpha_1+\alpha_2)}{2}}\beta_\rho\\&\bigg(J_{v_1-1}(\beta_\rho\rho)-J_{v_1+1}(\beta_\rho\rho)\bigg)\bigg)\times\bigg(\sin^{2-\alpha_2}\phi\cos^{2-\alpha_1}\phi\bigg(C_3 F(\alpha,\beta,\gamma;\zeta)+\\&C_4 \zeta^{1-\gamma}F(\alpha-\gamma+1,\beta-\gamma+1,2-\gamma;\zeta)\bigg)\bigg)\times\bigg(e^{kz}U(a,b;v_2)\bigg)\bigg]+\\&
        \bigg[i\frac{k}{2\eta}(\alpha_1+\alpha_2-2)\rho^{-\frac{(\alpha_1+\alpha_2)}{2}}\bigg(C_1 J_{v_1}(\beta_\rho\rho)\bigg)\times\bigg(\sin^{2-\alpha_2}\phi\cos^{2-\alpha_1}\phi\bigg(C_3\\& F(\alpha,\beta,\gamma;\zeta)+C_4\zeta^{1-\gamma}F(\alpha-\gamma+1,\beta-\gamma+1,2-\gamma;\zeta)\bigg)\bigg)\times\\&\bigg(e^{kz}U(a,b;v_2)\bigg)\bigg]
    \end{split}
\end{equation}

\begin{equation}\label{eq:D31}
    H_{ze}=0
\end{equation}
\end{widetext}
 
Similarly, $\rho$, $\phi$ and $z$ components of the TEn mode of the electric and magnetic fields can be calculated using Eq. \ref{eq:D12} and Eq. \ref{eq:D13}, as follows,
\begin{widetext}
\begin{equation}\label{eq:D32}
    \begin{split}
        E_{\rho m}={}&ik\eta\bigg[\frac{1}{\rho}\frac{\partial}{\partial \phi}\boldsymbol{\Pi_m}-\frac{\Pi_m}{2\rho}[(\alpha_1-1)\tan\phi-(\alpha_2-1)\cot\phi]\bigg]
    \end{split}
\end{equation}
\begin{equation}\label{eq:D33}
    \begin{split}
        E_{\rho m}={}&\bigg[ik\eta A_m\rho^{-\frac{(\alpha_1+\alpha_2)}{2}}\bigg(C_1 J_{v_1}(\beta_\rho\rho)\bigg)\times\bigg(\bigg((2-\alpha_2)\cos^{3-\alpha_1}\phi\sin^{1-\alpha_2}\phi\\&-(2-\alpha_1)\cos^{1-\alpha_1}\phi\sin^{3-\alpha_2}\phi\bigg)C_3F(\alpha,\beta,\gamma;\zeta)+2\sin^{3-\alpha_2}\phi\cos^{3-\alpha_1}\phi\\&\frac{\alpha\beta}{\gamma}C_3 F(\alpha+1,\beta+1,\gamma+1;\zeta)+\bigg((4-\alpha_2-2\gamma)\cos^{3-\alpha_1}\phi\\&\sin^{3-\alpha_2-2\gamma}\phi-(2-\alpha_1)\cos^{1-\alpha_1}\phi\sin^{5-\alpha_2-2\gamma}\phi\bigg)C_4\\& F(\alpha-\gamma+1,\beta-\gamma+1,2-\gamma;\zeta)+2\sin^{5-\alpha_2-2\gamma}\phi\cos^{3-\alpha_1}\phi\\&\frac{(\alpha-\gamma+1)(\beta-\gamma+1)}{(2-\gamma)}C_4 F(\alpha-\gamma+2,\beta-\gamma+2,3-\gamma;\zeta)\bigg)\times\\&\bigg(e^{kz}U(a,b;v_2)\bigg)\bigg]-
        \bigg[i\frac{k\eta}{2}A_m((\alpha_1-1)\tan\phi-(\alpha_2-1)\cot\phi)\\&\rho^{-\frac{(\alpha_1+\alpha_2)}{2}}\bigg(C_1 J_{v_1}(\beta_\rho\rho)\bigg)\times\bigg(\sin^{2-\alpha_2}\phi\cos^{2-\alpha_1}\phi\times\bigg(C_3 F(\alpha,\beta,\gamma;\zeta)+\\&C_4\zeta^{1-\gamma}F(\alpha-\gamma+1,\beta-\gamma+1,2-\gamma;\zeta)\bigg)\times\bigg(e^{kz}U(a,b;v_2)\bigg)\bigg]
    \end{split}
\end{equation}
\begin{equation}\label{eq:D34}
    \begin{split}
        E_{\phi m}={}&-ik\eta\bigg[\frac{\partial}{\partial\rho}\boldsymbol{\Pi_m}+\frac{\boldsymbol{\Pi_m}}{2\rho}(\alpha_1+\alpha_2-2)\bigg]
    \end{split}
\end{equation}
\begin{equation}\label{eq:D35}
    \begin{split}
        E_{\phi m}={}&\bigg[-ik\eta A_m\bigg(\bigg(1-\frac{(\alpha_1+\alpha_2)}{2}\bigg)\rho^{-\frac{(\alpha_1+\alpha_2)}{2}}C_1 J_{v_1}(\beta_\rho\rho)+\frac{C_1}{2}\rho^{1-\frac{(\alpha_1+\alpha_2)}{2}}\beta_\rho\\&\bigg(J_{v_1-1}(\beta_\rho\rho)-J_{v_1+1}(\beta_\rho\rho)\bigg)\bigg)\times\bigg(\sin^{2-\alpha_2}\phi\cos^{2-\alpha_1}\phi\bigg(C_3 F(\alpha,\beta,\gamma;\zeta)+\\&C_4 \zeta^{1-\gamma}F(\alpha-\gamma+1,\beta-\gamma+1,2-\gamma;\zeta)\bigg)\bigg)\times\bigg(e^{kz}U(a,b;v_2)\bigg)\bigg]-
        \bigg[i\frac{k\eta}{2}\\&A_m(\alpha_1+\alpha_2-2)\rho^{-\frac{(\alpha_1+\alpha_2)}{2}}\bigg(C_1 J_{v_1}(\beta_\rho\rho)\bigg)\times\bigg(\sin^{2-\alpha_2}\phi\cos^{2-\alpha_1}\phi\\&\bigg(C_3 F(\alpha,\beta,\gamma;\zeta)+C_4\zeta^{1-\gamma}F(\alpha-\gamma+1,\beta-\gamma+1,2-\gamma;\zeta)\bigg)\bigg)\times\\&\bigg(e^{kz}U(a,b;v_2)\bigg)\bigg]
    \end{split}
\end{equation}
\begin{equation}\label{eq:D36}
    E_{zm}=0
\end{equation}
\begin{equation}\label{eq:D37}
    \begin{split}
        H_{\rho m}{}&=[\boldsymbol{\nabla_D(\nabla_D.\Pi_m)}]_\rho=\bigg[\frac{\partial}{\partial\rho}+\frac{1}{2\rho}(\alpha_1+\alpha_2-2)\bigg](\boldsymbol{\nabla_D.\Pi_m})
    \end{split}
\end{equation}
\begin{equation}\label{eq:D38}
    \begin{split}
        H_{\rho m}={}&\bigg[A_m\bigg(\bigg(1-\frac{(\alpha_1+\alpha_2)}{2}\bigg)\rho^{-\frac{(\alpha_1+\alpha_2)}{2}}C_1(J_{v_1}(\beta_\rho\rho))+\frac{C1}{2}\rho^{1-\frac{(\alpha_1+\alpha_2)}{2}}\times\beta_\rho\\&(J_{v_1-1}(\beta_\rho\rho)-J_{v_1+1}(\beta_\rho\rho))\bigg)\times\bigg(\sin^{2-\alpha_2}\phi\cos^{2-\alpha_1}\phi\bigg(C_3F(\alpha,\beta,\gamma;\zeta)+\\&C_4\zeta^{1-\gamma}F(\alpha-\gamma+1,\beta-\gamma+1,2-\gamma;\zeta)\bigg)\bigg)\times\bigg(ke^{kz}U(a,b;v_2)-\\&ae^{kz}U(a+1,b+1;v_2)\bigg)\bigg]+ \bigg[\frac{(\alpha_3-1)}{2}A_m\bigg(\bigg(1-\frac{(\alpha_1+\alpha_2)}{2}\bigg)\rho^{-\frac{(\alpha_1+\alpha_2)}{2}}\\&C_1(J_{v_1}(\beta_\rho\rho))+\frac{C_1}{2}\rho^{1-\frac{(\alpha_1+\alpha_2)}{2}}\beta_\rho(J_{v_1-1}(\beta_\rho\rho)-J_{v_1+1}(\beta_\rho\rho))\bigg)\times\\&\bigg(\sin^{2-\alpha_2}\phi\cos^{2-\alpha_1}\phi\times(C_3F(\alpha,\beta,\gamma;\zeta)+C_4\zeta^{1-\gamma}\\&F(\alpha-\gamma+1,\beta-\gamma+1,2-\gamma;\zeta))\bigg)\times\bigg(z^{-1}e^{kz}U(a,b;v_2)\bigg)\bigg]+
        \\&\bigg[\frac{1}{2\rho}(\alpha_1+\alpha_2-2)(\boldsymbol{\nabla_D.\Pi_m})\bigg]
    \end{split}
\end{equation}
\begin{equation}\label{eq:D39}
    \begin{split}
        H_{\phi m}{}&=[\boldsymbol{\nabla_D(\nabla_D.\Pi_m)}]_{\phi}\\&=\bigg[\frac{1}{\rho}\frac{\partial}{\partial\phi}-\frac{1}{2\rho}\bigg((\alpha_1-1)\tan\phi-(\alpha_2-1)\cot\phi\bigg)\bigg](\boldsymbol{\nabla_D.\Pi_m})
    \end{split}
\end{equation}
\begin{equation}\label{eq:D40}
    \begin{split}
        H_{\phi m}={}&\bigg[A_m\rho^{-\frac{(\alpha_1+\alpha_2}{2}}\bigg(C_1J_{v_1}(\beta_\rho\rho)\bigg)\times\bigg(\bigg((2-\alpha_2)\cos^{3-\alpha_1}\phi\sin^{1-\alpha_2}\phi-\\&(2-\alpha_1)\cos^{1-\alpha_1}\phi\sin^{3-\alpha_2}\phi\bigg)C_3F(\alpha,\beta,\gamma;\zeta)+2\sin^{3-\alpha_2}\phi\cos^{3-\alpha_1}\phi\\&\frac{\alpha\beta}{\gamma}C_3F(\alpha+1,\beta+1,\gamma+1;\zeta)+\bigg((4-\alpha_2-2\gamma)\cos^{3-\alpha_1}\phi\\&\sin^{3-\alpha_2-2\gamma}\phi-(2-\alpha_1)\cos^{1-\alpha_1}\phi\sin^{5-\alpha_2-2\gamma}\phi\bigg)\\&C_4
        F(\alpha-\gamma+1,\beta-\gamma+1,2-\gamma;\zeta)+2C_4\\&F(\alpha-\gamma+2,\beta-\gamma+2,3-\gamma;\zeta)\sin^{5-\alpha_2-2\gamma}\phi\cos^{3-\alpha_1}\phi\times
        \\&\frac{(\alpha-\gamma+1)(\beta-\gamma+1)}{(2-\gamma)}\bigg)\times
        \bigg(ke^{kz}U(a,b;v_2)-ae^{kz}\\&U(a+1,b+1;v_2)\bigg)\bigg]+
        \bigg[\frac{(\alpha_3-1)}{2}A_m\rho^{-\frac{(\alpha_1+\alpha_2)}{2}}\bigg(C_1J_{v_1}(\beta_\rho\rho)\bigg)\times\\&\bigg(\bigg((2-\alpha_2)\cos^{3-\alpha_1}\phi\sin^{1-\alpha_2}\phi-(2-\alpha_1)\cos^{1-\alpha_1}\phi\sin^{3-\alpha_2}\phi\bigg)\\& C_3F(\alpha,\beta,\gamma;\zeta)+2\sin^{3-\alpha_2}\phi\cos^{3-\alpha_1}\phi\times\frac{\alpha\beta}{\gamma}\times\\& C_3F(\alpha+1,\beta+1,\gamma+1;\zeta)+\bigg((4-\alpha_2-2\gamma)\cos^{3-\alpha_1}\phi\sin^{3-\alpha_2-2\gamma}\phi-\\&(2-\alpha_1)\cos^{1-\alpha_1}\phi\sin^{5-\alpha_2-2\gamma}\phi\bigg)C_4F(\alpha-\gamma+1,\beta-\gamma+1,2-\gamma;\zeta)+\\&2\sin^{5-\alpha_2-2\gamma}\phi\cos^{3-\alpha_1}\phi\frac{(\alpha-\gamma+1)(\beta-\gamma+1)}{(2-\gamma)}C_4\\&F(\alpha-\gamma+2,\beta-\gamma+2,3-\gamma;\zeta))\bigg)\times\bigg(z^{-1}e^{kz}U(a,b;v_2)\bigg)\bigg]-\\&\bigg[\frac{1}{2\rho}((\alpha_1-1)\tan\phi-(\alpha_2-1)\cot\phi)(\boldsymbol{\nabla_D.\Pi_m})\bigg]
    \end{split}
\end{equation}
\begin{equation}\label{eq:D41}
    \begin{split}
       H_{zm}{}&=[\boldsymbol{\nabla_D(\nabla_D.\Pi_m)}]_z-k^2\Pi_m=\bigg[\frac{\partial}{\partial z}+\frac{(\alpha_3-1)}{2z}\bigg][\boldsymbol{\nabla_D.\Pi_m}]-k^2\boldsymbol{\Pi_m}
    \end{split}
\end{equation}
\begin{equation}\label{eq:D42}
    \begin{split}
        H_{zm}={}&\bigg[A_m\rho^{1-\frac{(\alpha_1+\alpha_2)}{2}}\bigg(C_1J_{v_1}(\beta_\rho\rho)\bigg)\times\bigg(\sin^{2-\alpha_2}\phi\cos^{2-\alpha_1}\phi\times
        \bigg(C_3\\&F(\alpha,\beta,\gamma;\zeta)+C_4\zeta^{1-\gamma}F(\alpha-\gamma+1,\beta-\gamma+1,2-\gamma;\zeta)\bigg)\times\\&\bigg(k^2e^{kz}U(a,b,v_2)-2ake^{kz}U(a+1,b+1;v_2)+a(a+1)e^{kz}\\&U(a+2,b+2;v_2)\bigg)\bigg]+\bigg[\frac{(\alpha_3-1)}{2}A_m\rho^{1-\frac{(\alpha_1+\alpha_2)}{2}}\bigg(C_1 J_{v_1}(\beta_\rho\rho)\bigg)\times\\&\bigg(C_3F(\alpha,\beta,\gamma;\zeta)+C_4
        \zeta^{1-\gamma}F(\alpha-\gamma+1,\beta-\gamma+1,2-\gamma;\zeta)\bigg)\times\\&\bigg(\frac{e^{kz}}{z^2}(kz-1)U(a,b;v_2)-\frac{ae^{kz}}{z}U(a+1,b+1;v_2)\bigg]+\\&\bigg[\frac{(\alpha_3-1)}{2z}(\boldsymbol{\nabla_D.\Pi_m})\bigg]-
        \bigg[k^2A_m\rho^{1-\frac{(\alpha_1+\alpha_2)}{2}}\bigg(C_1 J_{v_1}(\beta_\rho\rho)\bigg)\times\\&\bigg(\sin^{2-\alpha_2}\phi\cos^{2-\alpha_1}\phi\times\bigg(C_3F(\alpha,\beta,\gamma;\zeta)+C_4\zeta^{1-\gamma}\\&F(\alpha-\gamma+1,\beta-\gamma+1,2-\gamma;\zeta)\bigg)\bigg)\times\bigg(e^{kz}U(a,b;v_2)\bigg)\bigg]
    \end{split}
\end{equation}
\end{widetext}
 

\clearpage
\bibliographystyle{apsrev4-2}
\bibliography{main}

\providecommand{\noopsort}[1]{}\providecommand{\singleletter}[1]{#1}%
\begin{thebibliography}{48}%
\makeatletter
\providecommand \@ifxundefined [1]{%
 \@ifx{#1\undefined}
}%
\providecommand \@ifnum [1]{%
 \ifnum #1\expandafter \@firstoftwo
 \else \expandafter \@secondoftwo
 \fi
}%
\providecommand \@ifx [1]{%
 \ifx #1\expandafter \@firstoftwo
 \else \expandafter \@secondoftwo
 \fi
}%
\providecommand \natexlab [1]{#1}%
\providecommand \enquote  [1]{``#1''}%
\providecommand \bibnamefont  [1]{#1}%
\providecommand \bibfnamefont [1]{#1}%
\providecommand \citenamefont [1]{#1}%
\providecommand \href@noop [0]{\@secondoftwo}%
\providecommand \href [0]{\begingroup \@sanitize@url \@href}%
\providecommand \@href[1]{\@@startlink{#1}\@@href}%
\providecommand \@@href[1]{\endgroup#1\@@endlink}%
\providecommand \@sanitize@url [0]{\catcode `\\12\catcode `\$12\catcode
  `\&12\catcode `\#12\catcode `\^12\catcode `\_12\catcode `\%12\relax}%
\providecommand \@@startlink[1]{}%
\providecommand \@@endlink[0]{}%
\providecommand \url  [0]{\begingroup\@sanitize@url \@url }%
\providecommand \@url [1]{\endgroup\@href {#1}{\urlprefix }}%
\providecommand \urlprefix  [0]{URL }%
\providecommand \Eprint [0]{\href }%
\providecommand \doibase [0]{https://doi.org/}%
\providecommand \selectlanguage [0]{\@gobble}%
\providecommand \bibinfo  [0]{\@secondoftwo}%
\providecommand \bibfield  [0]{\@secondoftwo}%
\providecommand \translation [1]{[#1]}%
\providecommand \BibitemOpen [0]{}%
\providecommand \bibitemStop [0]{}%
\providecommand \bibitemNoStop [0]{.\EOS\space}%
\providecommand \EOS [0]{\spacefactor3000\relax}%
\providecommand \BibitemShut  [1]{\csname bibitem#1\endcsname}%
\let\auto@bib@innerbib\@empty
\bibitem [{\citenamefont {da~Silva}\ \emph {et~al.}(2020)\citenamefont
  {da~Silva}, \citenamefont {Pinillos}, \citenamefont {Tasca}, \citenamefont
  {Oxman},\ and\ \citenamefont {Khoury}}]{structured1}%
  \BibitemOpen
  \bibfield  {author} {\bibinfo {author} {\bibfnamefont {B.~P.}\ \bibnamefont
  {da~Silva}}, \bibinfo {author} {\bibfnamefont {V.}~\bibnamefont {Pinillos}},
  \bibinfo {author} {\bibfnamefont {D.}~\bibnamefont {Tasca}}, \bibinfo
  {author} {\bibfnamefont {L.}~\bibnamefont {Oxman}},\ and\ \bibinfo {author}
  {\bibfnamefont {A.}~\bibnamefont {Khoury}},\ }\href@noop {} {\bibfield
  {journal} {\bibinfo  {journal} {Physical review letters}\ }\textbf {\bibinfo
  {volume} {124}},\ \bibinfo {pages} {033902} (\bibinfo {year}
  {2020})}\BibitemShut {NoStop}%
\bibitem [{\citenamefont {Melo}\ \emph {et~al.}(2020)\citenamefont {Melo},
  \citenamefont {Brand\~ao}, \citenamefont {Pinheiro~da}, \citenamefont
  {Rodrigues}, \citenamefont {Khoury},\ and\ \citenamefont
  {Guerreiro}}]{structured2}%
  \BibitemOpen
  \bibfield  {author} {\bibinfo {author} {\bibfnamefont {B.}~\bibnamefont
  {Melo}}, \bibinfo {author} {\bibfnamefont {I.}~\bibnamefont {Brand\~ao}},
  \bibinfo {author} {\bibfnamefont {B.~S.}\ \bibnamefont {Pinheiro~da}},
  \bibinfo {author} {\bibfnamefont {R.}~\bibnamefont {Rodrigues}}, \bibinfo
  {author} {\bibfnamefont {A.}~\bibnamefont {Khoury}},\ and\ \bibinfo {author}
  {\bibfnamefont {T.}~\bibnamefont {Guerreiro}},\ }\href
  {https://doi.org/10.1103/PhysRevApplied.14.034069} {\bibfield  {journal}
  {\bibinfo  {journal} {Phys. Rev. Applied}\ }\textbf {\bibinfo {volume}
  {14}},\ \bibinfo {pages} {034069} (\bibinfo {year} {2020})}\BibitemShut
  {NoStop}%
\bibitem [{\citenamefont {Miller~Jr}(1977)}]{miller1977symmetry}%
  \BibitemOpen
  \bibfield  {author} {\bibinfo {author} {\bibfnamefont {W.}~\bibnamefont
  {Miller~Jr}},\ }\href@noop {} {\emph {\bibinfo {title} {Symmetry and
  separation of variables}}}\ (\bibinfo  {publisher} {Addison-Wesley Publishing
  Co., Inc., Reading, MA},\ \bibinfo {year} {1977})\BibitemShut {NoStop}%
\bibitem [{\citenamefont {{Brittingham}}(1982)}]{yu1}%
  \BibitemOpen
  \bibfield  {author} {\bibinfo {author} {\bibfnamefont {J.}~\bibnamefont
  {{Brittingham}}},\ }in\ \href {https://doi.org/10.1109/APS.1982.1148820}
  {\emph {\bibinfo {booktitle} {1982 Antennas and Propagation Society
  International Symposium}}},\ Vol.~\bibinfo {volume} {20}\ (\bibinfo {year}
  {1982})\ pp.\ \bibinfo {pages} {656--660}\BibitemShut {NoStop}%
\bibitem [{\citenamefont {Durnin}(1987)}]{yu5durnin1}%
  \BibitemOpen
  \bibfield  {author} {\bibinfo {author} {\bibfnamefont {J.}~\bibnamefont
  {Durnin}},\ }\href {https://doi.org/10.1364/JOSAA.4.000651} {\bibfield
  {journal} {\bibinfo  {journal} {J. Opt. Soc. Am. A}\ }\textbf {\bibinfo
  {volume} {4}},\ \bibinfo {pages} {651} (\bibinfo {year} {1987})}\BibitemShut
  {NoStop}%
\bibitem [{\citenamefont {Arlt}\ and\ \citenamefont {Dholakia}(2000)}]{yu7}%
  \BibitemOpen
  \bibfield  {author} {\bibinfo {author} {\bibfnamefont {J.}~\bibnamefont
  {Arlt}}\ and\ \bibinfo {author} {\bibfnamefont {K.}~\bibnamefont
  {Dholakia}},\ }\href {https://doi.org/10.1016/S0030-4018(00)00572-1}
  {\bibfield  {journal} {\bibinfo  {journal} {Optics Communications}\ }\textbf
  {\bibinfo {volume} {177}},\ \bibinfo {pages} {297} (\bibinfo {year}
  {2000})}\BibitemShut {NoStop}%
\bibitem [{\citenamefont {{Garces-Chavez}}\ \emph {et~al.}(2002)\citenamefont
  {{Garces-Chavez}}, \citenamefont {{Arlt}}, \citenamefont {{Dholakia}},
  \citenamefont {{Volke-Sepulveda}},\ and\ \citenamefont
  {{Chavez-Cerda}}}]{yu8}%
  \BibitemOpen
  \bibfield  {author} {\bibinfo {author} {\bibfnamefont {V.}~\bibnamefont
  {{Garces-Chavez}}}, \bibinfo {author} {\bibfnamefont {J.}~\bibnamefont
  {{Arlt}}}, \bibinfo {author} {\bibfnamefont {K.}~\bibnamefont {{Dholakia}}},
  \bibinfo {author} {\bibfnamefont {K.}~\bibnamefont {{Volke-Sepulveda}}},\
  and\ \bibinfo {author} {\bibfnamefont {S.}~\bibnamefont {{Chavez-Cerda}}},\
  }in\ \href {https://doi.org/10.1109/CLEO.2002.1033647} {\emph {\bibinfo
  {booktitle} {Summaries of Papers Presented at the Lasers and Electro-Optics.
  CLEO '02. Technical Diges}}}\ (\bibinfo {year} {2002})\ pp.\ \bibinfo {pages}
  {223--224 vol.1}\BibitemShut {NoStop}%
\bibitem [{\citenamefont {Arlt}\ \emph {et~al.}(2001)\citenamefont {Arlt},
  \citenamefont {Garces-Chavez}, \citenamefont {Sibbett},\ and\ \citenamefont
  {Dholakia}}]{opticalMicromanipulation}%
  \BibitemOpen
  \bibfield  {author} {\bibinfo {author} {\bibfnamefont {J.}~\bibnamefont
  {Arlt}}, \bibinfo {author} {\bibfnamefont {V.}~\bibnamefont {Garces-Chavez}},
  \bibinfo {author} {\bibfnamefont {W.}~\bibnamefont {Sibbett}},\ and\ \bibinfo
  {author} {\bibfnamefont {K.}~\bibnamefont {Dholakia}},\ }\href
  {https://doi.org/https://doi.org/10.1016/S0030-4018(01)01479-1} {\bibfield
  {journal} {\bibinfo  {journal} {Optics Communications}\ }\textbf {\bibinfo
  {volume} {197}},\ \bibinfo {pages} {239 } (\bibinfo {year}
  {2001})}\BibitemShut {NoStop}%
\bibitem [{\citenamefont {Cizmar}\ \emph {et~al.}(2004)\citenamefont {Cizmar},
  \citenamefont {Garces-Chavez}, \citenamefont {Dholakia},\ and\ \citenamefont
  {Zemanek}}]{OpticaltrappingCounter-propagating}%
  \BibitemOpen
  \bibfield  {author} {\bibinfo {author} {\bibfnamefont {T.}~\bibnamefont
  {Cizmar}}, \bibinfo {author} {\bibfnamefont {V.}~\bibnamefont
  {Garces-Chavez}}, \bibinfo {author} {\bibfnamefont {K.}~\bibnamefont
  {Dholakia}},\ and\ \bibinfo {author} {\bibfnamefont {P.}~\bibnamefont
  {Zemanek}},\ }in\ \href {https://doi.org/10.1117/12.557188} {\emph {\bibinfo
  {booktitle} {Optical Trapping and Optical Micromanipulation}}},\ Vol.\
  \bibinfo {volume} {5514},\ \bibinfo {editor} {edited by\ \bibinfo {editor}
  {\bibfnamefont {K.}~\bibnamefont {Dholakia}}\ and\ \bibinfo {editor}
  {\bibfnamefont {G.~C.}\ \bibnamefont {Spalding}}},\ \bibinfo {organization}
  {International Society for Optics and Photonics}\ (\bibinfo  {publisher}
  {SPIE},\ \bibinfo {year} {2004})\ pp.\ \bibinfo {pages} {643 --
  651}\BibitemShut {NoStop}%
\bibitem [{\citenamefont {Paterson}(2004)}]{LynnThesis}%
  \BibitemOpen
  \bibfield  {author} {\bibinfo {author} {\bibfnamefont {L.}~\bibnamefont
  {Paterson}},\ }\emph {\bibinfo {title} {Novel micromanipulation techniques in
  optical tweezers}},\ \href@noop {} {Ph.D. thesis},\ \bibinfo  {school}
  {University of St Andrews} (\bibinfo {year} {2004})\BibitemShut {NoStop}%
\bibitem [{\citenamefont {Li}\ and\ \citenamefont
  {Wang}(2017)}]{adaptiveFreeSpaceOpticalCoomunication}%
  \BibitemOpen
  \bibfield  {author} {\bibinfo {author} {\bibfnamefont {S.}~\bibnamefont
  {Li}}\ and\ \bibinfo {author} {\bibfnamefont {J.}~\bibnamefont {Wang}},\
  }\href {https://doi.org/10.1038/srep43233} {\bibfield  {journal} {\bibinfo
  {journal} {Scientific Reports}\ }\textbf {\bibinfo {volume} {7}},\ \bibinfo
  {pages} {43233} (\bibinfo {year} {2017})}\BibitemShut {NoStop}%
\bibitem [{\citenamefont {Wang}\ \emph {et~al.}(2012)\citenamefont {Wang},
  \citenamefont {Yang}, \citenamefont {Fazal}, \citenamefont {Ahmed},
  \citenamefont {Yan}, \citenamefont {Huang}, \citenamefont {Ren},
  \citenamefont {Yue}, \citenamefont {Dolinar}, \citenamefont {Tur},\ and\
  \citenamefont {Willner}}]{Terabitfree-space}%
  \BibitemOpen
  \bibfield  {author} {\bibinfo {author} {\bibfnamefont {J.}~\bibnamefont
  {Wang}}, \bibinfo {author} {\bibfnamefont {J.-Y.}\ \bibnamefont {Yang}},
  \bibinfo {author} {\bibfnamefont {I.}~\bibnamefont {Fazal}}, \bibinfo
  {author} {\bibfnamefont {N.}~\bibnamefont {Ahmed}}, \bibinfo {author}
  {\bibfnamefont {Y.}~\bibnamefont {Yan}}, \bibinfo {author} {\bibfnamefont
  {H.}~\bibnamefont {Huang}}, \bibinfo {author} {\bibfnamefont
  {Y.}~\bibnamefont {Ren}}, \bibinfo {author} {\bibfnamefont {Y.}~\bibnamefont
  {Yue}}, \bibinfo {author} {\bibfnamefont {S.}~\bibnamefont {Dolinar}},
  \bibinfo {author} {\bibfnamefont {M.}~\bibnamefont {Tur}},\ and\ \bibinfo
  {author} {\bibfnamefont {A.}~\bibnamefont {Willner}},\ }\href
  {https://doi.org/10.1038/nphoton.2012.138} {\bibfield  {journal} {\bibinfo
  {journal} {Nature Photonics}\ }\textbf {\bibinfo {volume} {6}},\ \bibinfo
  {pages} {488} (\bibinfo {year} {2012})}\BibitemShut {NoStop}%
\bibitem [{\citenamefont {Du}\ and\ \citenamefont {Wang}(2015)}]{Du:15}%
  \BibitemOpen
  \bibfield  {author} {\bibinfo {author} {\bibfnamefont {J.}~\bibnamefont
  {Du}}\ and\ \bibinfo {author} {\bibfnamefont {J.}~\bibnamefont {Wang}},\
  }\href {https://doi.org/10.1364/OL.40.004827} {\bibfield  {journal} {\bibinfo
   {journal} {Opt. Lett.}\ }\textbf {\bibinfo {volume} {40}},\ \bibinfo {pages}
  {4827} (\bibinfo {year} {2015})}\BibitemShut {NoStop}%
\bibitem [{\citenamefont {Mclaren}\ \emph {et~al.}(2014)\citenamefont
  {Mclaren}, \citenamefont {Mhlanga}, \citenamefont {Padgett}, \citenamefont
  {Roux},\ and\ \citenamefont {Forbes}}]{quantumEntanglement}%
  \BibitemOpen
  \bibfield  {author} {\bibinfo {author} {\bibfnamefont {M.}~\bibnamefont
  {Mclaren}}, \bibinfo {author} {\bibfnamefont {T.}~\bibnamefont {Mhlanga}},
  \bibinfo {author} {\bibfnamefont {M.}~\bibnamefont {Padgett}}, \bibinfo
  {author} {\bibfnamefont {F.}~\bibnamefont {Roux}},\ and\ \bibinfo {author}
  {\bibfnamefont {A.}~\bibnamefont {Forbes}},\ }\href
  {https://doi.org/10.1038/ncomms4248} {\bibfield  {journal} {\bibinfo
  {journal} {Nature communications}\ }\textbf {\bibinfo {volume} {5}},\
  \bibinfo {pages} {3248} (\bibinfo {year} {2014})}\BibitemShut {NoStop}%
\bibitem [{\citenamefont {Chen}\ \emph {et~al.}(2016)\citenamefont {Chen},
  \citenamefont {Khorasaninejad}, \citenamefont {Zhu}, \citenamefont {Oh},
  \citenamefont {Devlin}, \citenamefont {Zaidi},\ and\ \citenamefont
  {Capasso}}]{CapassoBessel}%
  \BibitemOpen
  \bibfield  {author} {\bibinfo {author} {\bibfnamefont {W.}~\bibnamefont
  {Chen}}, \bibinfo {author} {\bibfnamefont {M.}~\bibnamefont
  {Khorasaninejad}}, \bibinfo {author} {\bibfnamefont {A.}~\bibnamefont {Zhu}},
  \bibinfo {author} {\bibfnamefont {J.}~\bibnamefont {Oh}}, \bibinfo {author}
  {\bibfnamefont {R.}~\bibnamefont {Devlin}}, \bibinfo {author} {\bibfnamefont
  {A.}~\bibnamefont {Zaidi}},\ and\ \bibinfo {author} {\bibfnamefont
  {F.}~\bibnamefont {Capasso}},\ }\href {https://doi.org/10.1038/lsa.2016.259}
  {\bibfield  {journal} {\bibinfo  {journal} {Light: Science \& Applications}\
  }\textbf {\bibinfo {volume} {6}},\ \bibinfo {pages} {e16259} (\bibinfo {year}
  {2016})}\BibitemShut {NoStop}%
\bibitem [{\citenamefont {Yang}\ \emph {et~al.}(2018)\citenamefont {Yang},
  \citenamefont {Zhang}, \citenamefont {Bai},\ and\ \citenamefont
  {Wang}}]{YangSpiral:18}%
  \BibitemOpen
  \bibfield  {author} {\bibinfo {author} {\bibfnamefont {Z.}~\bibnamefont
  {Yang}}, \bibinfo {author} {\bibfnamefont {X.}~\bibnamefont {Zhang}},
  \bibinfo {author} {\bibfnamefont {C.}~\bibnamefont {Bai}},\ and\ \bibinfo
  {author} {\bibfnamefont {M.}~\bibnamefont {Wang}},\ }\href
  {https://doi.org/10.1364/JOSAA.35.000452} {\bibfield  {journal} {\bibinfo
  {journal} {J. Opt. Soc. Am. A}\ }\textbf {\bibinfo {volume} {35}},\ \bibinfo
  {pages} {452} (\bibinfo {year} {2018})}\BibitemShut {NoStop}%
\bibitem [{\citenamefont {{Zhang}}\ and\ \citenamefont
  {{Wang}}(2019)}]{wangbessel1}%
  \BibitemOpen
  \bibfield  {author} {\bibinfo {author} {\bibfnamefont {D.}~\bibnamefont
  {{Zhang}}}\ and\ \bibinfo {author} {\bibfnamefont {X.}~\bibnamefont
  {{Wang}}},\ }in\ \href {https://doi.org/10.1109/APUSNCURSINRSM.2019.8889035}
  {\emph {\bibinfo {booktitle} {2019 IEEE International Symposium on Antennas
  and Propagation and USNC-URSI Radio Science Meeting}}}\ (\bibinfo {year}
  {2019})\ pp.\ \bibinfo {pages} {437--438}\BibitemShut {NoStop}%
\bibitem [{\citenamefont {Yang}\ \emph {et~al.}(2019)\citenamefont {Yang},
  \citenamefont {Zhou},\ and\ \citenamefont {Wang}}]{wangbessel2}%
  \BibitemOpen
  \bibfield  {author} {\bibinfo {author} {\bibfnamefont {X.}~\bibnamefont
  {Yang}}, \bibinfo {author} {\bibfnamefont {Y.}~\bibnamefont {Zhou}},\ and\
  \bibinfo {author} {\bibfnamefont {G.}~\bibnamefont {Wang}},\ }\href
  {https://doi.org/10.1002/mmce.21941} {\bibfield  {journal} {\bibinfo
  {journal} {International Journal of RF and Microwave Computer-Aided
  Engineering}\ }\textbf {\bibinfo {volume} {29}},\ \bibinfo {pages} {e21941}
  (\bibinfo {year} {2019})}\BibitemShut {NoStop}%
\bibitem [{\citenamefont {Lin}\ \emph {et~al.}(2019)\citenamefont {Lin},
  \citenamefont {Li}, \citenamefont {Zhao}, \citenamefont {Song}, \citenamefont
  {Wang},\ and\ \citenamefont
  {Huang}}]{HighefficiencyBesselbeamarraygenerationbyHuygensmetasurfaces}%
  \BibitemOpen
  \bibfield  {author} {\bibinfo {author} {\bibfnamefont {Z.}~\bibnamefont
  {Lin}}, \bibinfo {author} {\bibfnamefont {X.}~\bibnamefont {Li}}, \bibinfo
  {author} {\bibfnamefont {R.}~\bibnamefont {Zhao}}, \bibinfo {author}
  {\bibfnamefont {X.}~\bibnamefont {Song}}, \bibinfo {author} {\bibfnamefont
  {Y.}~\bibnamefont {Wang}},\ and\ \bibinfo {author} {\bibfnamefont
  {L.}~\bibnamefont {Huang}},\ }\href@noop {} {\bibfield  {journal} {\bibinfo
  {journal} {Nanophotonics}\ }\textbf {\bibinfo {volume} {8}},\ \bibinfo
  {pages} {1079 } (\bibinfo {year} {2019})}\BibitemShut {NoStop}%
\bibitem [{\citenamefont {Akram}\ \emph {et~al.}(2019)\citenamefont {Akram},
  \citenamefont {Mehmood}, \citenamefont {Tauqeer}, \citenamefont {Rana},
  \citenamefont {Rukhlenko},\ and\ \citenamefont
  {Zhu}}]{highlyEfficientBessel}%
  \BibitemOpen
  \bibfield  {author} {\bibinfo {author} {\bibfnamefont {M.~R.}\ \bibnamefont
  {Akram}}, \bibinfo {author} {\bibfnamefont {M.~Q.}\ \bibnamefont {Mehmood}},
  \bibinfo {author} {\bibfnamefont {T.}~\bibnamefont {Tauqeer}}, \bibinfo
  {author} {\bibfnamefont {A.~S.}\ \bibnamefont {Rana}}, \bibinfo {author}
  {\bibfnamefont {I.~D.}\ \bibnamefont {Rukhlenko}},\ and\ \bibinfo {author}
  {\bibfnamefont {W.}~\bibnamefont {Zhu}},\ }\href
  {https://doi.org/10.1364/OE.27.009467} {\bibfield  {journal} {\bibinfo
  {journal} {Opt. Express}\ }\textbf {\bibinfo {volume} {27}},\ \bibinfo
  {pages} {9467} (\bibinfo {year} {2019})}\BibitemShut {NoStop}%
\bibitem [{\citenamefont {Dudley}\ \emph {et~al.}(2013)\citenamefont {Dudley},
  \citenamefont {Lavery}, \citenamefont {Padgett},\ and\ \citenamefont
  {Forbes}}]{unravelingbesselbeams}%
  \BibitemOpen
  \bibfield  {author} {\bibinfo {author} {\bibfnamefont {A.}~\bibnamefont
  {Dudley}}, \bibinfo {author} {\bibfnamefont {M.}~\bibnamefont {Lavery}},
  \bibinfo {author} {\bibfnamefont {M.}~\bibnamefont {Padgett}},\ and\ \bibinfo
  {author} {\bibfnamefont {A.}~\bibnamefont {Forbes}},\ }\href
  {https://doi.org/10.1364/OPN.24.6.000022} {\bibfield  {journal} {\bibinfo
  {journal} {Opt. Photon. News}\ }\textbf {\bibinfo {volume} {24}},\ \bibinfo
  {pages} {22} (\bibinfo {year} {2013})}\BibitemShut {NoStop}%
\bibitem [{\citenamefont {Tao}\ and\ \citenamefont {Yuan}(2004)}]{Tao1}%
  \BibitemOpen
  \bibfield  {author} {\bibinfo {author} {\bibfnamefont {S.~H.}\ \bibnamefont
  {Tao}}\ and\ \bibinfo {author} {\bibfnamefont {X.}~\bibnamefont {Yuan}},\
  }\href {https://doi.org/10.1364/JOSAA.21.001192} {\bibfield  {journal}
  {\bibinfo  {journal} {J. Opt. Soc. Am. A}\ }\textbf {\bibinfo {volume}
  {21}},\ \bibinfo {pages} {1192} (\bibinfo {year} {2004})}\BibitemShut
  {NoStop}%
\bibitem [{\citenamefont {Tao}\ \emph {et~al.}(2003)\citenamefont {Tao},
  \citenamefont {Lee},\ and\ \citenamefont {Yuan}}]{Tao2}%
  \BibitemOpen
  \bibfield  {author} {\bibinfo {author} {\bibfnamefont {S.~H.}\ \bibnamefont
  {Tao}}, \bibinfo {author} {\bibfnamefont {W.~M.}\ \bibnamefont {Lee}},\ and\
  \bibinfo {author} {\bibfnamefont {X.-C.}\ \bibnamefont {Yuan}},\ }\href
  {https://doi.org/10.1364/OL.28.001867} {\bibfield  {journal} {\bibinfo
  {journal} {Opt. Lett.}\ }\textbf {\bibinfo {volume} {28}},\ \bibinfo {pages}
  {1867} (\bibinfo {year} {2003})}\BibitemShut {NoStop}%
\bibitem [{\citenamefont {Tao}\ \emph {et~al.}(2004)\citenamefont {Tao},
  \citenamefont {Lee},\ and\ \citenamefont {Yuan}}]{Tao3}%
  \BibitemOpen
  \bibfield  {author} {\bibinfo {author} {\bibfnamefont {S.~H.}\ \bibnamefont
  {Tao}}, \bibinfo {author} {\bibfnamefont {W.~M.}\ \bibnamefont {Lee}},\ and\
  \bibinfo {author} {\bibfnamefont {X.}~\bibnamefont {Yuan}},\ }\href
  {https://doi.org/10.1364/AO.43.000122} {\bibfield  {journal} {\bibinfo
  {journal} {Appl. Opt.}\ }\textbf {\bibinfo {volume} {43}},\ \bibinfo {pages}
  {122} (\bibinfo {year} {2004})}\BibitemShut {NoStop}%
\bibitem [{\citenamefont {Tao}\ \emph {et~al.}(2005)\citenamefont {Tao},
  \citenamefont {Yuan}, \citenamefont {Lin}, \citenamefont {Peng},\ and\
  \citenamefont {Niu}}]{Tao4}%
  \BibitemOpen
  \bibfield  {author} {\bibinfo {author} {\bibfnamefont {S.~H.}\ \bibnamefont
  {Tao}}, \bibinfo {author} {\bibfnamefont {X.-C.}\ \bibnamefont {Yuan}},
  \bibinfo {author} {\bibfnamefont {J.}~\bibnamefont {Lin}}, \bibinfo {author}
  {\bibfnamefont {X.}~\bibnamefont {Peng}},\ and\ \bibinfo {author}
  {\bibfnamefont {H.~B.}\ \bibnamefont {Niu}},\ }\href
  {https://doi.org/10.1364/OPEX.13.007726} {\bibfield  {journal} {\bibinfo
  {journal} {Opt. Express}\ }\textbf {\bibinfo {volume} {13}},\ \bibinfo
  {pages} {7726} (\bibinfo {year} {2005})}\BibitemShut {NoStop}%
\bibitem [{\citenamefont {Marston}(2009)}]{TaoComment}%
  \BibitemOpen
  \bibfield  {author} {\bibinfo {author} {\bibfnamefont {P.~L.}\ \bibnamefont
  {Marston}},\ }\href {https://doi.org/10.1364/JOSAA.26.002181} {\bibfield
  {journal} {\bibinfo  {journal} {J. Opt. Soc. Am. A}\ }\textbf {\bibinfo
  {volume} {26}},\ \bibinfo {pages} {2181} (\bibinfo {year}
  {2009})}\BibitemShut {NoStop}%
\bibitem [{\citenamefont {Gutierrez~Vega}\ and\ \citenamefont
  {López-Mariscal}(2007)}]{Vega1}%
  \BibitemOpen
  \bibfield  {author} {\bibinfo {author} {\bibfnamefont {J.}~\bibnamefont
  {Gutierrez~Vega}}\ and\ \bibinfo {author} {\bibfnamefont {C.}~\bibnamefont
  {López-Mariscal}},\ }\href {https://doi.org/10.1088/1464-4258/10/01/015009}
  {\bibfield  {journal} {\bibinfo  {journal} {Journal of Optics A: Pure and
  Applied Optics}\ }\textbf {\bibinfo {volume} {10}},\ \bibinfo {pages}
  {015009} (\bibinfo {year} {2007})}\BibitemShut {NoStop}%
\bibitem [{\citenamefont {L\'{o}pez-Mariscal}\ \emph
  {et~al.}(2008)\citenamefont {L\'{o}pez-Mariscal}, \citenamefont {Burnham},
  \citenamefont {Rudd}, \citenamefont {McGloin},\ and\ \citenamefont
  {Guti\'{e}rrez-Vega}}]{Vega2}%
  \BibitemOpen
  \bibfield  {author} {\bibinfo {author} {\bibfnamefont {C.}~\bibnamefont
  {L\'{o}pez-Mariscal}}, \bibinfo {author} {\bibfnamefont {D.}~\bibnamefont
  {Burnham}}, \bibinfo {author} {\bibfnamefont {D.}~\bibnamefont {Rudd}},
  \bibinfo {author} {\bibfnamefont {D.}~\bibnamefont {McGloin}},\ and\ \bibinfo
  {author} {\bibfnamefont {J.~C.}\ \bibnamefont {Guti\'{e}rrez-Vega}},\ }\href
  {https://doi.org/10.1364/OE.16.011411} {\bibfield  {journal} {\bibinfo
  {journal} {Opt. Express}\ }\textbf {\bibinfo {volume} {16}},\ \bibinfo
  {pages} {11411} (\bibinfo {year} {2008})}\BibitemShut {NoStop}%
\bibitem [{\citenamefont {Mitri}(2011)}]{Mitri1}%
  \BibitemOpen
  \bibfield  {author} {\bibinfo {author} {\bibfnamefont {F.~G.}\ \bibnamefont
  {Mitri}},\ }\href {https://doi.org/10.1364/OL.36.000606} {\bibfield
  {journal} {\bibinfo  {journal} {Opt. Lett.}\ }\textbf {\bibinfo {volume}
  {36}},\ \bibinfo {pages} {606} (\bibinfo {year} {2011})}\BibitemShut
  {NoStop}%
\bibitem [{\citenamefont {Mitri}(2012)}]{Mitri2}%
  \BibitemOpen
  \bibfield  {author} {\bibinfo {author} {\bibfnamefont {F.~G.}\ \bibnamefont
  {Mitri}},\ }\href {https://doi.org/10.1103/PhysRevA.85.025801} {\bibfield
  {journal} {\bibinfo  {journal} {Phys. Rev. A}\ }\textbf {\bibinfo {volume}
  {85}},\ \bibinfo {pages} {025801} (\bibinfo {year} {2012})}\BibitemShut
  {NoStop}%
\bibitem [{\citenamefont {Mitri}(2013)}]{Mitri3}%
  \BibitemOpen
  \bibfield  {author} {\bibinfo {author} {\bibfnamefont {F.~G.}\ \bibnamefont
  {Mitri}},\ }\href@noop {} {\bibfield  {journal} {\bibinfo  {journal} {The
  European Physical Journal D}\ }\textbf {\bibinfo {volume} {67}},\ \bibinfo
  {pages} {1} (\bibinfo {year} {2013})}\BibitemShut {NoStop}%
\bibitem [{\citenamefont {{Céspedes}}\ and\ \citenamefont
  {{Caloz}}(2019)}]{caloz}%
  \BibitemOpen
  \bibfield  {author} {\bibinfo {author} {\bibfnamefont {O.~V.}\ \bibnamefont
  {{Céspedes}}}\ and\ \bibinfo {author} {\bibfnamefont {C.}~\bibnamefont
  {{Caloz}}},\ }in\ \href
  {https://doi.org/10.1109/PIERS-Spring46901.2019.9017541} {\emph {\bibinfo
  {booktitle} {2019 PhotonIcs Electromagnetics Research Symposium - Spring
  (PIERS-Spring)}}}\ (\bibinfo {year} {2019})\ pp.\ \bibinfo {pages}
  {714--717}\BibitemShut {NoStop}%
\bibitem [{\citenamefont {Zubair}\ \emph
  {et~al.}(2012{\natexlab{a}})\citenamefont {Zubair}, \citenamefont {Mughal},\
  and\ \citenamefont {Naqvi}}]{ZubairBook}%
  \BibitemOpen
  \bibfield  {author} {\bibinfo {author} {\bibfnamefont {M.}~\bibnamefont
  {Zubair}}, \bibinfo {author} {\bibfnamefont {M.}~\bibnamefont {Mughal}},\
  and\ \bibinfo {author} {\bibfnamefont {Q.}~\bibnamefont {Naqvi}},\ }\href
  {https://doi.org/10.1007/978-3-642-25358-4} {\emph {\bibinfo {title}
  {Electromagnetic Fields and Waves in Fractional Dimensional Space}}}\
  (\bibinfo  {publisher} {Springer-Verlag Berlin Heidelberg},\ \bibinfo {year}
  {2012})\BibitemShut {NoStop}%
\bibitem [{\citenamefont {Zubair}\ \emph
  {et~al.}(2012{\natexlab{b}})\citenamefont {Zubair}, \citenamefont {Mughal},\
  and\ \citenamefont {Naqvi}}]{ZubairWavePropagation}%
  \BibitemOpen
  \bibfield  {author} {\bibinfo {author} {\bibfnamefont {M.}~\bibnamefont
  {Zubair}}, \bibinfo {author} {\bibfnamefont {M.~J.}\ \bibnamefont {Mughal}},\
  and\ \bibinfo {author} {\bibfnamefont {Q.~A.}\ \bibnamefont {Naqvi}},\ }in\
  \href@noop {} {\emph {\bibinfo {booktitle} {Electromagnetic fields and waves
  in fractional dimensional space}}}\ (\bibinfo  {publisher} {Springer},\
  \bibinfo {year} {2012})\ pp.\ \bibinfo {pages} {27--60}\BibitemShut {NoStop}%
\bibitem [{\citenamefont {Zubair}\ \emph
  {et~al.}(2011{\natexlab{a}})\citenamefont {Zubair}, \citenamefont {Mughal},\
  and\ \citenamefont {Naqvi}}]{ZubairWaveEquation}%
  \BibitemOpen
  \bibfield  {author} {\bibinfo {author} {\bibfnamefont {M.}~\bibnamefont
  {Zubair}}, \bibinfo {author} {\bibfnamefont {M.}~\bibnamefont {Mughal}},\
  and\ \bibinfo {author} {\bibfnamefont {Q.}~\bibnamefont {Naqvi}},\ }\href
  {https://doi.org/10.2528/PIERL10102103} {\bibfield  {journal} {\bibinfo
  {journal} {Progress In Electromagnetics Research Letters}\ }\textbf {\bibinfo
  {volume} {19}},\ \bibinfo {pages} {137} (\bibinfo {year}
  {2011}{\natexlab{a}})}\BibitemShut {NoStop}%
\bibitem [{\citenamefont {Zubair}\ \emph
  {et~al.}(2018{\natexlab{a}})\citenamefont {Zubair}, \citenamefont {Ang},\
  and\ \citenamefont {Ang}}]{ZubairNovelFractional}%
  \BibitemOpen
  \bibfield  {author} {\bibinfo {author} {\bibfnamefont {M.}~\bibnamefont
  {Zubair}}, \bibinfo {author} {\bibfnamefont {Y.~S.}\ \bibnamefont {Ang}},\
  and\ \bibinfo {author} {\bibfnamefont {L.~K.}\ \bibnamefont {Ang}},\
  }\href@noop {} {\bibfield  {journal} {\bibinfo  {journal} {2018 Progress in
  Electromagnetics Research Symposium (PIERS-Toyama)}\ ,\ \bibinfo {pages}
  {2533}} (\bibinfo {year} {2018}{\natexlab{a}})}\BibitemShut {NoStop}%
\bibitem [{\citenamefont {Zubair}\ \emph
  {et~al.}(2018{\natexlab{b}})\citenamefont {Zubair}, \citenamefont {Ang},
  \citenamefont {Ooi},\ and\ \citenamefont {Ang}}]{zubair4}%
  \BibitemOpen
  \bibfield  {author} {\bibinfo {author} {\bibfnamefont {M.}~\bibnamefont
  {Zubair}}, \bibinfo {author} {\bibfnamefont {Y.~S.}\ \bibnamefont {Ang}},
  \bibinfo {author} {\bibfnamefont {K.}~\bibnamefont {Ooi}},\ and\ \bibinfo
  {author} {\bibfnamefont {L.}~\bibnamefont {Ang}},\ }\href
  {https://doi.org/10.1063/1.5039811} {\bibfield  {journal} {\bibinfo
  {journal} {Journal of Applied Physics}\ }\textbf {\bibinfo {volume} {124}},\
  \bibinfo {pages} {163101} (\bibinfo {year} {2018}{\natexlab{b}})}\BibitemShut
  {NoStop}%
\bibitem [{\citenamefont {Zubair}\ \emph
  {et~al.}(2018{\natexlab{c}})\citenamefont {Zubair}, \citenamefont {Ang},\
  and\ \citenamefont {Ang}}]{zubair5}%
  \BibitemOpen
  \bibfield  {author} {\bibinfo {author} {\bibfnamefont {M.}~\bibnamefont
  {Zubair}}, \bibinfo {author} {\bibfnamefont {Y.~S.}\ \bibnamefont {Ang}},\
  and\ \bibinfo {author} {\bibfnamefont {L.~K.}\ \bibnamefont {Ang}},\
  }\href@noop {} {\bibfield  {journal} {\bibinfo  {journal} {IEEE Transactions
  on Electron Devices}\ }\textbf {\bibinfo {volume} {65}},\ \bibinfo {pages}
  {3421} (\bibinfo {year} {2018}{\natexlab{c}})}\BibitemShut {NoStop}%
\bibitem [{\citenamefont {Naqvi}\ and\ \citenamefont
  {Zubair}(2016)}]{zubair14}%
  \BibitemOpen
  \bibfield  {author} {\bibinfo {author} {\bibfnamefont {Q.}~\bibnamefont
  {Naqvi}}\ and\ \bibinfo {author} {\bibfnamefont {M.}~\bibnamefont {Zubair}},\
  }\href@noop {} {\bibfield  {journal} {\bibinfo  {journal} {Optik}\ }\textbf
  {\bibinfo {volume} {127}},\ \bibinfo {pages} {3243} (\bibinfo {year}
  {2016})}\BibitemShut {NoStop}%
\bibitem [{\citenamefont {Zubair}\ and\ \citenamefont {Ang}(2016)}]{zubair12}%
  \BibitemOpen
  \bibfield  {author} {\bibinfo {author} {\bibfnamefont {M.}~\bibnamefont
  {Zubair}}\ and\ \bibinfo {author} {\bibfnamefont {L.}~\bibnamefont {Ang}},\
  }\href@noop {} {\bibfield  {journal} {\bibinfo  {journal} {Physics of
  Plasmas}\ }\textbf {\bibinfo {volume} {23}},\ \bibinfo {pages} {072118}
  (\bibinfo {year} {2016})}\BibitemShut {NoStop}%
\bibitem [{\citenamefont {Zubair}\ \emph
  {et~al.}(2011{\natexlab{b}})\citenamefont {Zubair}, \citenamefont {Mughal},\
  and\ \citenamefont {Naqvi}}]{exactCylindrical}%
  \BibitemOpen
  \bibfield  {author} {\bibinfo {author} {\bibfnamefont {M.}~\bibnamefont
  {Zubair}}, \bibinfo {author} {\bibfnamefont {M.}~\bibnamefont {Mughal}},\
  and\ \bibinfo {author} {\bibfnamefont {Q.}~\bibnamefont {Naqvi}},\ }\href
  {https://doi.org/10.2528/PIER11021508} {\bibfield  {journal} {\bibinfo
  {journal} {Progress In Electromagnetics Research}\ }\textbf {\bibinfo
  {volume} {114}},\ \bibinfo {pages} {443} (\bibinfo {year}
  {2011}{\natexlab{b}})}\BibitemShut {NoStop}%
\bibitem [{\citenamefont {Kak}(2020)}]{frac_dimension}%
  \BibitemOpen
  \bibfield  {author} {\bibinfo {author} {\bibfnamefont {S.}~\bibnamefont
  {Kak}},\ }\href@noop {} {\bibfield  {journal} {\bibinfo  {journal}
  {Scientific Reports}\ }\textbf {\bibinfo {volume} {10}},\ \bibinfo {pages}
  {1} (\bibinfo {year} {2020})}\BibitemShut {NoStop}%
\bibitem [{\citenamefont {{Ehsan}}\ \emph {et~al.}(2020)\citenamefont
  {{Ehsan}}, \citenamefont {{Mehmood}}, \citenamefont {{Ang}}, \citenamefont
  {{Ang}},\ and\ \citenamefont {{Zubair}}}]{SFBB}%
  \BibitemOpen
  \bibfield  {author} {\bibinfo {author} {\bibfnamefont {A.}~\bibnamefont
  {{Ehsan}}}, \bibinfo {author} {\bibfnamefont {M.~Q.}\ \bibnamefont
  {{Mehmood}}}, \bibinfo {author} {\bibfnamefont {Y.~S.}\ \bibnamefont
  {{Ang}}}, \bibinfo {author} {\bibfnamefont {L.~K.}\ \bibnamefont {{Ang}}},\
  and\ \bibinfo {author} {\bibfnamefont {M.}~\bibnamefont {{Zubair}}},\ }in\
  \href@noop {} {\emph {\bibinfo {booktitle} {2020 14th European Conference on
  Antennas and Propagation (EuCAP)}}}\ (\bibinfo {year} {2020})\ pp.\ \bibinfo
  {pages} {1--5}\BibitemShut {NoStop}%
\bibitem [{\citenamefont {Li}\ \emph {et~al.}(2020)\citenamefont {Li},
  \citenamefont {Bongiovanni}, \citenamefont {Goutsoulas}, \citenamefont {Xia},
  \citenamefont {Zhang}, \citenamefont {Hu}, \citenamefont {Song},
  \citenamefont {Morandotti}, \citenamefont {Efremidis},\ and\ \citenamefont
  {Chen}}]{SLM1}%
  \BibitemOpen
  \bibfield  {author} {\bibinfo {author} {\bibfnamefont {D.}~\bibnamefont
  {Li}}, \bibinfo {author} {\bibfnamefont {D.}~\bibnamefont {Bongiovanni}},
  \bibinfo {author} {\bibfnamefont {M.}~\bibnamefont {Goutsoulas}}, \bibinfo
  {author} {\bibfnamefont {S.}~\bibnamefont {Xia}}, \bibinfo {author}
  {\bibfnamefont {Z.}~\bibnamefont {Zhang}}, \bibinfo {author} {\bibfnamefont
  {Y.}~\bibnamefont {Hu}}, \bibinfo {author} {\bibfnamefont {D.}~\bibnamefont
  {Song}}, \bibinfo {author} {\bibfnamefont {R.}~\bibnamefont {Morandotti}},
  \bibinfo {author} {\bibfnamefont {N.~K.}\ \bibnamefont {Efremidis}},\ and\
  \bibinfo {author} {\bibfnamefont {Z.}~\bibnamefont {Chen}},\ }\href
  {https://doi.org/10.1364/OSAC.391878} {\bibfield  {journal} {\bibinfo
  {journal} {OSA Continuum}\ }\textbf {\bibinfo {volume} {3}},\ \bibinfo
  {pages} {1525} (\bibinfo {year} {2020})}\BibitemShut {NoStop}%
\bibitem [{\citenamefont {Hu}\ \emph {et~al.}(2020)\citenamefont {Hu},
  \citenamefont {Tai}, \citenamefont {Zhu}, \citenamefont {Long}, \citenamefont
  {Tang}, \citenamefont {Li}, \citenamefont {Li},\ and\ \citenamefont
  {Cai}}]{SLM2}%
  \BibitemOpen
  \bibfield  {author} {\bibinfo {author} {\bibfnamefont {J.}~\bibnamefont
  {Hu}}, \bibinfo {author} {\bibfnamefont {Y.}~\bibnamefont {Tai}}, \bibinfo
  {author} {\bibfnamefont {L.}~\bibnamefont {Zhu}}, \bibinfo {author}
  {\bibfnamefont {Z.}~\bibnamefont {Long}}, \bibinfo {author} {\bibfnamefont
  {M.}~\bibnamefont {Tang}}, \bibinfo {author} {\bibfnamefont {H.}~\bibnamefont
  {Li}}, \bibinfo {author} {\bibfnamefont {X.}~\bibnamefont {Li}},\ and\
  \bibinfo {author} {\bibfnamefont {Y.}~\bibnamefont {Cai}},\ }\href
  {https://doi.org/10.1063/5.0004692} {\bibfield  {journal} {\bibinfo
  {journal} {Applied Physics Letters}\ }\textbf {\bibinfo {volume} {116}},\
  \bibinfo {pages} {201107} (\bibinfo {year} {2020})},\ \Eprint
  {https://arxiv.org/abs/https://doi.org/10.1063/5.0004692}
  {https://doi.org/10.1063/5.0004692} \BibitemShut {NoStop}%
\bibitem [{\citenamefont {Lu}\ \emph {et~al.}(2020)\citenamefont {Lu},
  \citenamefont {Cao}, \citenamefont {Zhu},\ and\ \citenamefont {Gu}}]{SLM3}%
  \BibitemOpen
  \bibfield  {author} {\bibinfo {author} {\bibfnamefont {J.}~\bibnamefont
  {Lu}}, \bibinfo {author} {\bibfnamefont {C.}~\bibnamefont {Cao}}, \bibinfo
  {author} {\bibfnamefont {Z.}~\bibnamefont {Zhu}},\ and\ \bibinfo {author}
  {\bibfnamefont {B.}~\bibnamefont {Gu}},\ }\href
  {https://doi.org/10.1063/5.0002756} {\bibfield  {journal} {\bibinfo
  {journal} {Applied Physics Letters}\ }\textbf {\bibinfo {volume} {116}},\
  \bibinfo {pages} {201105} (\bibinfo {year} {2020})},\ \Eprint
  {https://arxiv.org/abs/https://doi.org/10.1063/5.0002756}
  {https://doi.org/10.1063/5.0002756} \BibitemShut {NoStop}%
\bibitem [{\citenamefont {Abramowitz}(1974)}]{Abramowitz}%
  \BibitemOpen
  \bibfield  {author} {\bibinfo {author} {\bibfnamefont {M.}~\bibnamefont
  {Abramowitz}},\ }\href@noop {} {\emph {\bibinfo {title} {Handbook of
  Mathematical Functions, With Formulas, Graphs, and Mathematical Tables,}}}\
  (\bibinfo  {publisher} {Dover Publications, Inc.},\ \bibinfo {address} {New
  York, NY, USA},\ \bibinfo {year} {1974})\BibitemShut {NoStop}%
\bibitem [{\citenamefont {Stillinger}(1977)}]{Stillinger}%
  \BibitemOpen
  \bibfield  {author} {\bibinfo {author} {\bibfnamefont {F.~H.}\ \bibnamefont
  {Stillinger}},\ }\href {https://doi.org/10.1063/1.523395} {\bibfield
  {journal} {\bibinfo  {journal} {Journal of Mathematical Physics}\ }\textbf
  {\bibinfo {volume} {18}},\ \bibinfo {pages} {1224} (\bibinfo {year}
  {1977})}\BibitemShut {NoStop}%
\end{thebibliography}%

\end{document}